\documentclass[manuscript]{aastex}

\newcommand{\ee}[1]{\mbox{${} \times 10^{#1}$}}
\newcommand{\eten}[1]{\mbox{$10^{#1}$}}
\newcommand{\etal}{{\it et~al.}}

\newcommand{\degree}{\mbox{$^{\circ}$}}
\newcommand{\amin}{\mbox{\arcmin}}
\newcommand{\asec}{\mbox{$^{\prime\prime}$}}

\newcommand{\kms}{\mbox{km\,s$^{-1}$}}
\newcommand\cmv{\mbox{cm$^{-3}$}}

\newcommand{\um}{$\mu$m}
\newcommand{\wno}{cm$^{-1}$}

\def\lsim {$\rlap{\raise.4ex\hbox{$<$}}\lower.55ex\hbox{$\sim$}\,$}

\newcommand\fir{far-infrared}

\newcommand{\lsun}{\mbox{L$_\odot$}}
\newcommand{\msun}{\mbox{M$_\odot$}}

\newcommand{\neii}{[Ne\,II]}

\newcommand{\sgras}{Sgr~A$^*$}
\newcommand{\sgraw}{Sgr~A West}

\slugcomment{to be published in Astrophys. J.}

\shorttitle{Galactic Center Ionized Gas}
\shortauthors{Irons \etal}

\begin{document}


\title{Ionized Gas in the Galactic Center:
New Observations and Interpretation}


\author{Wesley T. Irons, John H. Lacy\altaffilmark{1}}
\affil{Department of Astronomy, University of Texas, Austin, TX 78712}
\email{lacy@astro.as.utexas.edu}

\and

\author{Matthew J. Richter\altaffilmark{1}}
\affil{Department of Physics, University of California, Davis, CA 95616}


\altaffiltext{1}{Visiting Astronomer at the Infrared Telescope Facility,
which is operated by the University of Hawaii under Cooperative Agreement no.
NCC 5-538 with the National Aeronautics and Space Administration, Office of
Space Science, Planetary Astronomy Program.}


\begin{abstract}
We present new observations of the \neii\ emission from the ionized gas in
\sgraw\ with improved resolution and sensitivity.
About half of the emission comes from gas with kinematics indicating it is
orbiting in a plane tipped about 25\degree\ from the Galactic plane.
This plane is consistent with that derived previously for the circumnuclear
molecular disk and the northern arm and western arc ionized features.
However, unlike most previous studies, we conclude that the ionized gas
is not moving along the ionized features, but on more nearly circular paths.
The observed speeds are close to, but probably somewhat less than expected
for orbital motions in the potential of the central black hole and stars
and have a small inward component.
The spatial distribution of the emission is well fitted by a spiral pattern.
We discuss possible physical explanations for the spatial distribution and
kinematics of the ionized gas, and conclude that both may be best explained
by a one-armed spiral density wave,
which also accounts for both the observed low velocities and the inward
velocity component.
We suggest that a density wave may result from the precession of elliptical
orbits in the potential of the black hole and stellar mass distribution.
\end{abstract}





\section{INTRODUCTION}

The center of the Milky Way Galaxy has been the subject of intense study
since the observation of infrared emission from the central star cluster
\citep{bn68} and radio wavelength emission from the ionized gas
in \sgraw\ \citep{downes71}
and the central compact object \sgras\ \citep{balick74}.
Being nearly 100 times closer than any other major galactic nucleus,
our Galactic center provides the best opportunity to observe the
interaction of stars and gas with a super-massive black hole (SMBH).
Numerous authors have reviewed the contents and phenomena found in
the Galactic center.
\citet{morris96}, \citet{mezger96}, and \citet{genzel10} discuss
observations of the interstellar gas that is most relevant for
this paper.

Observations of stellar proper motions and radial velocities
\citep{ghez08, gillessen09} give a distance to the Galactic center
of 8.3\,kpc (corresponding to an image scale of 25\asec /pc)
and a black hole mass of 4.3\ee{6}\,\msun.
Stellar spectra and imaging give evidence for recent star formation
or the capture of a recently formed star cluster \citep{levin03, gerhard01}.
The stellar mass distribution in \sgraw\ is not well known.
If there is an equilibrium stellar cusp, as expected for a cluster around
a SMBH \citep{bahcall76}, the radial dependence 
of the stellar density can be described by a broken power law with
a slope $\gamma \approx 1.3$ for the cusp and $\gamma \approx 1.8$
outside the cusp \citep{genzel03, schodel07, genzel10}.
\citet{merritt10} describes how the absence of a Bahcall-Wolf cusp is
plausible assuming that the mass in the inner parsec is traced by old stars
which would indicate a low-density core with radius $\approx$0.5\,pc.
Some recent papers \citep{buchholz09, do09, bartko10} have suggested
that there may be a relatively flat stellar density inside of $\sim$1~pc.

The most prominent interstellar matter in the central 1.5\,pc is
the ionized gas and the associated warm dust \citep{rieke88}.
This gas has the appearance of a clumpy, filamentary multi-armed
spiral \citep{lo83,serabyn85}.
The mass of the ionized gas is several tens of \msun.
Neutral atomic gas is also present in the inner few pc \citep{jackson93}.
Although it is more difficult to observe, its mass is a factor $\sim$10
times that of the ionized gas.
Beyond $\sim$1.5~pc and extending out to $\sim$10~pc,
the interstellar gas is mostly molecular, and is referred to as the
circumnuclear disk or CND
\citep{becklin82,gusten87,christopher05,montero09,oka11}.
Estimates for the mass of the CND range from a few \eten{4}\,\msun,
based on millimeter dust emission \citep{mezger89,davidson92,etxaluze11},
to \eten{6}\,\msun, based on virial masses of molecular clumps
\citep{christopher05,montero09}.

Observations of infrared and radio hydrogen recombination lines (RRLs)
\citep{roberts93,herbst93,paumard03,zhao09} and infrared fine-structure lines
\citep{wollman77,lacy80,serabyn88,lacy91} provide information on the
motion of the ionized gas through Doppler shifts.
The overall pattern is consistent with expectations for orbital
motions in a potential dominated by the massive black hole:
the highest velocities are found within a few arcseconds of \sgras,
and velocities tend to decrease going outward.
Much of the gas appears to be near a plane tipped $\sim 25^{\circ}$
from the Galactic
plane, with redshifts seen toward positive Galactic longitudes
and blueshifts generally seen toward negative longitudes.
The motions of the molecular gas in the CND are also mostly
in the sense of Galactic rotation, but with a roughly flat rotation
curve, as distributed mass makes a larger contribution to the
gravitational potential farther from the center.

Several models have been proposed to explain the gas kinematics.
\citet{lacy80} originally saw the gas as being in a number of
independently orbiting clouds, but better imaging,
especially with the VLA \citep{lo83}, showed that the ionized gas
was better described as a collection of streamers, with the `clouds'
being peaks in the emission along the streamers.
\citet{serabyn85} and \citet{serabyn88} found that Doppler shifts
vary smoothly along the streamers.
They modeled the `western arc' (see Fig. 1) as the ionized inner
rim of the CND in a nearly circular orbit around the center,
and the `northern arm' as a flow of gas approaching the center.
\citet{lacy91} obtained a complete data cube of the \neii\ emission
from the inner 60\asec\,$\times$\,90\asec\ and
concluded that the gas kinematics of the western arc and northern arm
were better modeled with circular motions, rather than motions along
the streamers.  They argued that the western arc and northern arm
are orbiting in the same plane as the CND and that they could be
joined at their north ends to form a single spiral feature.
The main problem with their interpretation was the lack of a
physical explanation for the spiral.
They suggested that it could be a density wave or a spiraling
inflow affected by both gravitational and viscous forces, but in
both cases it was hard to identify the forces responsible for
organizing the gas into a spiral pattern.
Observations of infrared and radio hydrogen recombination line emission
led various authors \citep{sanders98,vollmer00,liszt03,paumard04}
to return to the tidally stretched cloud model.
\citet{zhao09} strengthened this model by including proper motions
of the ionized gas streamers.
They fitted observations of the western arc, northern arm, and eastern
arm with elliptical Keplerian orbits in the potential of the central
black hole.

Non-gravitational forces may also influence the gas distribution
and motions.  \citet{aitken91,aitken98} and \citet{glasse03}
observed polarized emission
from the dust in the northern arm and bar region, indicating that
mGauss magnetic fields are aligned along the ionized streamers.
\citet{aitken98} interpreted variations in the polarization
to give a measure of the inclination of the magnetic fields from the
plane of the sky.
Assuming that the flows are along the field lines, they obtained
information about the 3-dimensional structure of the gas orbits.
Stellar winds apparently also affect the ionized gas.
In several cases bow shocks are seen around stars, presumably as
the stars move through the ionized medium or a wind from the
central region blows past the stars \citep{serabyn91, geballe04}.

We have made new observations of the \neii\ emission from \sgraw\
with improved spectral and spatial resolution, as well as improved
sensitivity.
In this paper, we present these observations and compare them to the
different models of the ionized gas kinematics.

\section{OBSERVATIONS}

\neii\ line emission from \sgraw\ was observed in June 2003
with TEXES on the NASA IRTF.
TEXES, the Texas Echelon Cross Echelle Spectrograph \citep{lacy02},
is a high resolution spectrograph for wavelengths of 5-25\,\um.
At the 12.8\,\um\ (780.42\,\wno ) wavelength of the \neii\ fine-structure line
it has a spectral resolution of $\sim$.01\,\wno, or 4\,\kms,
and a spatial resolution along the entrance slit of $\sim$1.3\asec.
Spectral and spatial sampling is 4 pixels per resolution element.
This is a significant improvement over that of the data cube used
by \citet{lacy91}, which had a spectral resolution of 33\,\kms
and a spatial resolution of $\sim$2\asec.
For comparison, the observations presented by \citet{zhao09}
have a spectral resolution of 14\,\kms\ (with an HI thermal linewidth
$\sim$20\,\kms ) and a spatial resolution of $\sim$1.25\asec.
To map \sgraw, the spectrograph slit was oriented N-S and stepped 
in 0.7\asec\ steps to the east, covering 76\asec.
As the slit length is only 10\asec, multiple scans were made,
separated by 5\asec\ in declination, to cover a region
of $76 \times 78$\asec\ centered near \sgras.
At 12.8\,\um\ the echelon spectral orders are about 10\% longer than
the detector width, leaving gaps between orders, with $\sim$230\,\kms\
of each 255\,\kms\ order covered.
To fill in these gaps, we made two sets of observations:
one with the spectrum offset to center the red side of each echelon
order on the detector array and one with the blue side centered.
The spectral coverage of the combined data cube is -1300\,--\,800\,\kms.

The individual scans were reduced as described by \citet{lacy02},
and the sky emission was subtracted from each using the median
value along the scan as sky for each spectral pixel.
We also subtracted fluctuating water vapor emission by subtracting
a multiple of a spectrum obtained from the correlation of each
spectral pixel with pixels containing strong water lines.
The scans were merged to make two large maps with the spectral
settings described above.
The merging procedure involved cross-correlating contour maps of the 
scans to determine the appropriate offsets.
The two maps were then combined, averaging overlapping spectral pixels,
to complete the data cube.
Absolute coordinates were obtained by aligning ionized gas peaks
with peaks in the VLA continuum maps.
Due to uncertainties in the procedure to merge scans into a map,
absolute and relative coordinates have uncertainties $\sim$1\arcsec.

The data cube was deconvolved using a maximum entropy method (MEM)
deconvolution routine, utilizing an algorithm from \citet{nityananda82}.
This routine is designed to sharpen the image when the signal is high, 
while smoothing low signal regions.  
The routine effectively smoothed out the noise while slightly enhancing
the resolution in regions where the line is apparent. 
The contour map of the data cube, summed over Doppler shifts
from $-339$\,\kms\ to  $+299$\,\kms\ is shown in Figure 1,
with the various filaments labeled.

A FITS format deconvolved data cube, spectrally binned by two pixels
and including -670\,\kms\ to +345\,\kms\ is available in the on-line journal.

\section{KINEMATIC FITTING}

Using the new Galactic center observations of \neii\ emission,
we compare two models for the ionized gas motions in the region.
The first models the filaments as tidally stretched clouds in which
the gas is flowing along the streamers. 
For this model, we use the fitting parameters of \citet{zhao09}
where the gas filaments are taken to be on separate Keplerian orbits.
We refer to this as the ``ellipse model."
This model is the most widely accepted explanation for the ionized gas
kinematics.
The second model we discuss is similar to the model of \citet{lacy91}.
This model fits the gas on nearly circular orbits around \sgras\ and
connects the northern arm and the western arc into a single feature
in one plane.
We will refer to this model as the ``circular orbit model," although
we will allow the orbits to have a nonzero radial component. 
The circular orbit model has not received as much attention, 
but we are motivated to reevaluate it with the improved data 
and some discrepancies between the data and the ellipse model.

\subsection{\it Ellipse Model}

The orbits fitted by \citet{zhao09} are shown on the \neii\ image in Figure 2.
To compare our data to this model, we first smoothed the \neii\ data cube
spatially over a square box on the sky of width [17\,R$_{pc}$ pixels],
where R$_{pc}$ is the radial distance from the center in parsecs.
The data cube was then tagged with points along the ellipses separated by
1$^{\circ}$ in the orbital planes, and spectra were extracted from the
tagged points and used to construct the position-velocity diagrams shown
in Figure 3.
The model curves in Figure 3 assume Keplerian motions along the ellipses
and a SMBH mass of 4.2\ee{6}\,\msun.

The elliptical orbit models fit rather well spatially, with emission seen
typically for $\sim$180\degree\ along each orbit.
The spectral fit, however, shows some discrepancies from the data.
For the northern arm (Fig. 3a) the data are systematically offset 
in velocity from the model.
This would be explained if the velocities in the northern arm are not
aligned with the streamer as assumed, but are systematically tipped toward
circular motions.
This anomaly supports the previous assumption by \citet{lacy91} that the
gas is moving in circular orbits across the filaments.
There is a similar anomaly in Figure 3b, though less pronounced, 
as the proposed orbit for the western arc is almost circular.

We note that Figure 20 of \citet{zhao09} shows a similar discrepancy
between the elliptical orbit model and the RRL data.
In general, the \neii\ data agree with the RRL data,
giving us confidence that the \neii\ line is a good tracer of the 
ionized gas.

The kinematic model for the eastern arm (Figure 3c) seems to be the
least convincing.
Unfortunately, the circular orbit model also does not fit
the eastern arm kinematics, so it is clear that more work is needed
to accurately describe that filament.
\citet{liszt03} proposes a combined model for the eastern arm and the bar
that deserves further study.

The disagreement between the elliptical orbit models and the kinematic
data indicates that the gas does not move along the ionized streamers,
especially the northern arm.
But before discarding this type of model, we should ask whether a
tidally stretched cloud is necessarily stretched along its direction
of motion.
To answer this question we ran a simple two-dimensional hydrodynamic
model, similar to that discussed in section 5.2.
In fact, whether the particles in an infalling cloud follow along a
narrow streamer depends on the initial conditions of the simulation.
In particular, if the cloud initially rotates in the prograde direction
it forms an arc as it approaches the center, with the particle motions
tipped from the arc toward more circular paths.
Although this suggests that a tidally stretched cloud model could be
made which would agree better with the observations than those we
considered, we were not able to find a very acceptable model of this type.
Consequently, we now proceed to consider other types of models.

\subsection{\it Circular Orbit Model}

The discrepancies in the ellipse models provide motivation to reevaluate the 
circular orbit model of \citet{lacy91} with the improved TEXES data.
Assuming circular velocities, we fit the \neii\ emission spectrally with 
parameters describing the plane in which the gas is moving
(that is, an inclination angle, $\imath$, and the position angle of the line of nodes,
$\Omega$), along with those describing the mass distribution in the region.
The orbital speed was taken to be $v = [G M(r) / r]^{1/2}$,
appropriate for orbital motions in a spherically symmetric mass
distribution.
If non-gravitational forces act on the gas or if the mass distribution
is not spherically symmetric, $M(r)$ can be taken to be
a way of parameterizing the forces acting on the gas.
The best fit was that plane which fit the most gas from the entire
data cube on circular orbits, using a procedure described below.
We also considered models in which we included a correction to the
circular Doppler velocity by adding
an inward (or outward) radial component that allows the gas to be
spiraling into (or out of) the Galactic center. 
This radial component was made a free parameter, 
and could take the value zero.

For our calculations, we used two different stellar mass distributions. 
The first is a power law distribution adopted from 
\citet{genzel10}\footnote{The distribution determined by \citet{genzel10}
was a broken power law.  We are using a slightly simplified form of
their mass distribution for $r>0.25\,pc$ noting that our observations are
not particularly sensitive to the stellar mass and that these numbers contain
large systematic uncertainties.}
\begin{equation}
M_{*}(r) \approx M_{1} (\frac{r}{1.0pc})^{\alpha}.
\end{equation}
Here, we did not allow M$_{1}$ and the power  $\alpha$
to be free parameters as the ionized gas is not especially sensitive
to the stellar mass distribution, so they will take the values
$1.0\ee{6}\,\msun$ and 1.2, respectively.
The second mass distribution considered assumes a Lorentzian density
distribution
\begin{equation}
{\rho} (r) \approx {\rho}_{\circ} (\frac{R_{c}^{2}}{r^{2}+R_{c}^{2}}),
\end{equation}
where ${\rho}_{\circ}$ is the density at the center and $R_{c}$
is the core radius. 
This distribution flattens out within the core radius and drops as $1/r^{2}$ 
outside the core radius, roughly approximating an isothermal star cluster.
Equation 2 can be integrated to arrive at a mass distribution
\begin{equation}
M_{*} (r) \approx 4 {\pi} {\rho}_{\circ} R_{c}^{2} (r - R_{c} \arctan (\frac{r}{R_{c}})).
\end{equation}

For each mass distribution model, we determined the best fit allowing
the dynamical mass of \sgras\ and the core stellar density, as well as
the disk plane orientation, to be free parameters.
Because the black hole mass is well known from stellar observations 
\citep{ghez08, gillessen09} and the \neii\ data are not especially sensitive
to the stellar mass, it was useful to also find the best fits holding these
parameters constant ($M_{\bullet}=4.2\ee{6}\,\msun$, 
$\rho_{\circ}=1\ee{6}\,\msun\,pc^{-3}$, $R_{c}=0.5$\,pc).

To determine the best parameters for the orbiting gas, we first
searched through parameter space to find the best-fit orbital plane.
For each set of parameters describing a plane and orbital motion,
we calculated the Doppler velocity at each spatial point in the data cube.
We then shifted the spectrum at each point by that velocity 
so that all the emission fitting circular velocities in that plane would
be aligned at the zero velocity wavelength,
and collapsed the map into a single spectrum.
We then compared how well the emission in the data cube fit
the model velocities in a given plane with any other plane.
To illustrate what this routine does we show in Figure 4
the collapsed spectrum before the spectral shifting (4a) and
the spectrum corresponding to a good fit (4b).
Of course this method will always provide a ``best fit" set of parameters
for any velocity model, and
it is necessary to determine how good such a fit is.
We have no quantitative way of stating error bars for the fit parameters
because some velocity variations are expected due to turbulence,
and the discrepancies from the fit are probably not due to random
noise with any known distribution.
In addition, we do not expect our model to fit all of the emission,
as we know that not all of the gas lies in a single plane.
After smoothing the shifted spectrum with a Gaussian, we can determine
the flux at zero Doppler velocity, and compare the results with each
set of parameters.
Alternatively, we can compare the flux in an interval around zero
Doppler velocity.
The fitting procedure utilized the first method with a smoothing
Gaussian defined by
$\exp [-\Delta p^{2}/\sigma^{2}]$, where $\Delta p$
is the separation from the spectral pixel to the zero velocity pixel
and $\sigma=5$\,\kms.
We also give the results using the second method using a $\pm$30\,\kms\
interval in Table 1.
For a good fit, we would expect to gather a higher percentage of the total
\neii\ emission in a lower percentage of the total velocity range.
We provide these ratios as an easy way to interpret the results of the fitting,
though this can be done essentially by looking at the Gaussian fit to the
collapsed spectrum that results from this routine.

In order to allow the gas to spiral into or out of the Galactic center,
we added to the calculated circular velocities a radial component 
which is equal to the fraction $a/r$ of the angular component, 
where $a$ is a free parameter.
The gas can then be viewed as flowing along a linear spiral,
$r_{flow} (\theta) = a \theta$,
where $\theta$ is measured in radians.
The collapsed spectrum in Figure 4b was made using the mass distribution 
in Equation 3 and the parameters, $\imath = 66^{\circ}$, $\Omega = 23^{\circ}$,
M$_{\bullet} = 3.5\ee{6}\,\msun$, $\rho_{\circ} = 2.5\ee{5}\,\msun$\,pc$^{-3}$, 
R$_{c} = 0.5$\,pc, and $a = -0.06$\,pc.

The results of this fitting are compiled in Table 1.
It includes the best fits for both stellar mass distributions,
Equation 1 and Equation 3,
holding the black hole mass fixed and allowing it to change.
We also determined the best fit using HCN(4-3) data from \citet{montero09},
to determine whether the orbital plane and mass distribution that fit the
ionized gas also fit the molecular gas in the CND.
We also quote a ``goodness parameter" which is just the ratio of the emission
that fits within $\pm$\,30\,\kms\ to the total emission in the map.
Again, this is not the same as quoting error bars, but it provides a rough
measure of whether this fit could have been achieved by accident.

To accompany Table 1 and the ``goodness parameter," Figure 5 shows 
the collapsed, shifted spectra corresponding to the different fits in the table.
Each set of conditions in Table 1 provides an acceptable best fit,
as seen in Figure 5. 
The ionized gas is not a good measure of the shape of the mass distribution
and whether the mass should be in the black hole or in the stars.
With the different mass distributions and mass constraints, 
the best fits all give roughly the same orbital plane,
with the exception of row 8, which is described below.

Note that when using Equation 3 for the stellar mass distribution, 
the best fit black hole mass is smaller than
the accepted values from the literature.
Alternatively, if we constrain the mass to be 4.2\ee{6}\,\msun,
we prefer no stars in the region.
Although there clearly are stars in \sgraw, 
this result may be consistent with the suggestion that the stellar density
is flat within $\sim$1~pc, with a possible hole in the inner few arcseconds
\citep{buchholz09, do09}.
We will speculate further on the low mass results in Section 5, but for now
we will just examine the results shown in the fifth row of Table 1.

In Figure 6a we show the spatial distribution of the emission within
$\pm$\,30\,\kms\ of the velocity of the best-fitting disk model.
This emission gives a strong impression of a spiral pattern,
as noted by \citet{lacy91}, and (partly to lead the eye) we superimpose
on the image a nearly linear (Archimedean) spiral in the disk plane
given by $r(\theta)=0.27$\,pc\,$\times \theta(rad)^{0.93}$.
Note that the disk plane used is that derived from the gas kinematics,
not the spatial pattern, but it allows a good fit to the pattern.
To define the spatial distribution for the spiral, we must specify
another parameter which we take as the third Euler angle, $\phi$,
(the others being $\imath$ and $\Omega$).
The angle $\phi$ describes the starting point of the spiral.
Its effect is the same as adding a constant to $r(\theta)$ and letting $\theta$ go negative.
In Figure 6a, $\phi = 274^{\circ}$.
We note that our kinematic model involves gas moving along almost circular
orbits, not along the spiral, implying that gas moves across the streamers.
For the kinematic model presented here we use the parameters 
from the fifth row of Table 1.
The disk parameters for the spiral structure are similar to those used by 
\citet{lacy91} and by \citet{zhao09} for the western arc, with 
an inclination angle $\imath = 66^{\circ}$ and an
angle of line of nodes $\Omega = 23^{\circ}$.
We display in Figure 6b the integrated \neii\ emission with the emission shown
in Figure 6a masked out to see the gas that does not lie in the fitted plane.
Comparing to Figure 1, we can see from Figures 6a and 6b that most of the 
northern arm and western arc emission fits the velocity pattern described above,
while the eastern arm and bar, as well as diffuse emission do not.
Note that the integrated intensity in Figure 6a is 45\% of that
for the whole map (Figure 1). 
This is the basis of the ``goodness of fit" for our method.
It strengthens the argument that the fitting routine is not merely
collecting random gas coincidentally if the percentage of the total emission
that fits within a certain velocity range is
greater than the percentage of the total velocity range in which it fits.
So, in this case, the $\pm$\,30\,\kms\ range is less than 10\% of the total
velocity range for the map, while the emission collected in this range is
about 45\% of the total emission.

To determine how well the velocities of the gas match circular orbits
as opposed to motions along the features we compare the position-velocity
(P-V) diagrams for gas along the spiral shown in Figure 6a, using
the velocity patterns for purely circular orbits (row 5 in Table 1,
but with $a=0$\,pc) and for motion along the spiral
(row 5 in Table 1, but with $a=-0.27$\,pc). 
These two models are superposed on the P-V diagram in Figure 7a.
The best fit parameters include a small inward radial component to the 
velocity ($a=-0.06$\,pc),
which is included in the P-V diagram in Figure 7b.
For the purely circular velocity pattern (lower curve in Figure 7a), 
there is a slight offset in position angle along the northern arm,
which validates
the small nonzero inward radial velocity component in the best fit.
The best fit parameters, including the radial velocity correction, 
make an excellent fit to the data (Figure 7b).
We also tried to apply an inward radial velocity component with a 
constant pitch angle, so the radial component would be $c\,v_{\theta}$
rather than $(a/r)v_{\theta}$.
Here, the parameter $c$ just shifts the P-V diagram up
or down depending on the sign.
The best fit in this case had a much smaller pitch angle, which had a 
negligible effect on the P-V diagram.
This is because when the model is shifted up to agree with the data along
the northern arm, the model along the western arc no longer fits.
Therefore, spiraling motion described by $v_{r}=(a/r)v_{\theta}$
fits the data best.
The top curve in Figure 7a is for motion along the physical spiral 
in the best fit plane. 
We note that the model is offset from the data in the same way
the ellipse model was offset.
This is further confirmation that the gas is not flowing along the streamers.

Although the best kinematic fit to gas near the northern arm / western
arc plane involved little inward motion, we wanted to determine whether
motion along a spiral in a different plane could fit the kinematic data
while providing an acceptable fit to the spatial distribution.
There was not a good fit that worked both spatially and spectrally.
The best fit spectrally (for motion along a spiral) corresponded to a 
very poor fit spatially.
By allowing the plane orientation parameter to vary substantially, 
we were able to fit the kinematics with a larger value of $a$ ($-0.1$\,pc),
but the spiral described by this value of $a$ does not follow the 
observed spatial distribution.

\subsection{{\it Molecular Emission in the Circumnuclear Disk}}

We show in Figure 8 a contour map of HCN(4-3) emission from 
data by \citet{montero09} superposed on the grayscale image of the
\neii\ emission.
We used a procedure like that used with the \neii\ data to find the
best-fitting plane and mass distribution to explain the HCN kinematics.
Much of the gas in the circumnuclear disk has been found to fit in a plane 
with inclination, $\imath \approx 70^{\circ}$ \citep{jackson93}.
We can see from rows 3 and 7 of Table 1 that our results are similar and
that this plane is near the plane that fits the \neii\ kinematics.
Note that the CND data are relatively insensitive to the inclination
parameter as the rotation curve is mostly flat.
When fitting the HCN(4-3) emission, we constrained
the black hole mass, as the gas motions in the CND are more sensitive 
to the stellar mass.
The resulting stellar mass parameters are consistent with those derived 
in row 5 of Table 1.
However, the inward radial velocity component, $a$, is consistent with
motions along the spiral that fits the ionized gas observations.
At $R = 2$\,pc, in the CND, $a = -0.4$\,pc corresponds to a pitch angle
of $a/R=0.2$ and an inward velocity of 20\,\kms.

The western arc has been described as the ionized inner rim of the CND,
so it is interesting to ask if the spiral pattern continues into the
molecular gas.
The HCN(4-3) emission that fits circular velocities in the plane of the spiral
within $\pm$30\,\kms~ is shown in Figure 9. 
Included on the HCN image is the spiral from the \neii\ map in Figure 6a
(tagged every 1$^{\circ}$), and the same spiral offset in starting position 
by 75$^{\circ}$ (tagged every 3$^{\circ}$). 
Much of the HCN(4-3) emission lies just outside the spiral used for the
ionized gas, or along the second spiral.
Figure 10 shows the HCN P-V diagram extracted from
along the outside spiral from Figure 9.

The model in Figure 10 does not fit as well as we had hoped, perhaps because
the random motions in the CND are a larger fraction of the orbital speed
than those in the ionized spiral, though it does fit acceptably well,
particularly in the region southwest of \sgras.

\section{CONCLUSIONS FROM THE OBSERVATIONS}

Before discussing theoretical models and implications from our observations,
we state the conclusions we have drawn that are independent of those
models.

Approximately half of the ionic line emission from \sgraw\ comes from
gas orbiting in a plane tipped about 25\degree\ from the Galactic plane.
This plane is coincident within uncertainties with that of the
molecular circumnuclear disk.

The gas in the disk plane moves on nearly circular orbits, with only
a small inward velocity component.
The Doppler pattern is not consistent with motion along the northern
arm ionized streamer.

The observed speeds are close to, but probably somewhat less
than expected for orbital motions in the gravitational potential of
the central super-massive black hole and the distributed mass,
as derived from the orbital motions of stars near the black hole
and the distribution of stars.

The spatial distribution of the ionized gas in the western arc and
northern arm could be described by two ellipses in a plane close to
that derived from the [Ne~II] line kinematics, but is somewhat better
fitted with a single, approximately linear (Archimedean) spiral.

\section{DISCUSSION AND INTERPRETATION}

The fact that the Doppler shifts of the ionized gas in the northern
arm and western arc require nearly circular motions rather than
motions along the streamers has been pointed out before \citep{lacy91}.
The elliptical model has become prevalent, perhaps mostly because
the interpretation of these features as tidally stretched clouds is
so natural, and because it is hard to understand how the ionized gas
could be concentrated in an eccentric streamer like the northern arm
if the gas motions are circular.
We now add to this puzzle about the direction of motion of the gas
the observation that the gas speed is probably less than expected in
the potential of the black hole and star cluster.
Both of these observations need to be explained.

\subsection{\it Non-gravitational Forces}

We first ask about the possible significance
of the fact that the best fit to the kinematic data involves orbital
motions somewhat less than the expected Keplerian velocities.
One explanation for sub-Keplerian velocities is that there are
non-gravitational outward forces on the ionized gas,
which might be caused by radiation pressure or by ram pressure of
a hot wind.
These forces might also contribute to the organization of the gas
into the observed spiral pattern.

The radiation pressure due to Thomson scattering of photons from
the central star cluster off of electrons can easily be shown to be
negligible, as the electron scattering opacity is very small.
The dust opacity is larger, so radiation pressure on dust mixed
with the ionized gas should be considered.
The importance of radiation pressure can be estimated by comparing
the momentum flux from a stellar luminosity of a few \eten{6}\,\lsun\ to
the gravitational force on a parcel of gas with a column density
corresponding to an optical depth to starlight of order one,
assuming a normal interstellar dust to gas ratio.
This calculation indicates that radiation pressure is a factor
$\sim$100 less important than gravity, so probably not enough to
account for the sub-Keplerian velocities.

If we assume that the x-ray emitting hot gas in the region generates
a ram pressure
equal to the thermal pressure of the gas, i.e. that the gas moves at
the sound speed, we can estimate the ratio of the outward
force from the hot gas to the gravitational force.
Using reasonable estimates for the size of a gas cloud and the hot gas
temperature and electron density from \citet{baganoff03},
that ratio is about \eten{-3}\, to \eten{-2}.
As an outward force, this will not cause the low velocities observed, 
but if it is seen as a drag force as the ionized gas flows past the hot gas,
it might be enough to cause the slight spiraling inward motion 
observed in the ionized gas.

Rather than explaining the possible sub-Keplerian velocities with
an outward force, we could explain them by hypothesizing that the
gas slows its orbital motion due to a shock as it enters the spiral,
and possibly the dissipation of kinetic energy is at least partially
responsible for the ionization of the gas.
A 20\% decrease in the gas speed would be consistent with the
observed Doppler shifts and would correspond to a 40\% loss in kinetic energy.
However, even if all of the kinetic energy of the gas entering the northern
arm were converted to ionization, the ionization rate would not balance
the recombination rate derived from the free-free observation.
So although a shock might contribute to the ionization of
the gas, and it may affect the orbital speed of the gas,
it is not likely to dominate the ionization.

Magnetic forces are likely to be greater than other non-gravitational forces.
We can estimate their importance by comparing the magnetic energy density,
or pressure, with the kinetic energy density of the orbiting gas and the
gas pressure.
\citet{aitken98} give a lower limit on the field strength of 2\,mG.
This corresponds to an energy density of
$B^2 / {8\pi} = 1.6\times10^{-7}$\,erg\,\cmv.
If we estimate the ionized gas density to be \eten{4}\,cm$^{-3}$
(based on our unpublished observations of [S~III], which has a
critical density of 1.7\ee{4}), a mean particle mass of \eten{-24}\,g,
an orbital speed of 100\,\kms , and a temperature of 8000\,K,
we calculate a kinetic energy density of 5\ee{-6}\,erg\,\cmv\ 
and a gas pressure of 8\ee{-9}\,erg\,\cmv.
Apparently the magnetic pressure is substantially greater than the
gas pressure, but probably substantially smaller than the kinetic
energy density.
Consequently, we would expect the field to be carried along by
the gas without altering the orbital speeds significantly,
although it may exert forces which could perturb the gas
motions, possibly influencing the density wave discussed below.

\subsection {\it A Spiral Density Wave} 

We now turn to the origin of the spiral pattern and the observation
that the gas moves across the northern arm and western arc streamers.
There are two main problems with the idea that the gas motions do not
align with the streamers.
First, if the streamers move with the gas, they would wrap up quickly.
The orbital period varies from a few $10^3$ years at the inner end
of the northern arm, $\sim$\,0.1 pc from the center, to $\sim 10^5$ yr
at the inner edge of the CND, so that a wrap should be added to the
spiral pattern every few $10^3$ yr.
On the other hand, if the gas moves across the streamers we need to
explain why the emission is concentrated there.
We also should explain how ionized gas could move across magnetic
field lines, which the observations of \citet{aitken98} clearly show
to be aligned with the northern arm.

In the case of galactic spiral arms, the wrapping problem is normally
resolved by assuming that the gas and stars move through the arms,
and are observed to be concentrated there because their orbits crowd
and spend more time in the arms.
Perhaps the ionized gas spiral in the Galactic center is also caused
by a density wave.
This possibility is supported by the fact that a one-armed spiral is
the dominant instability in a disk with a nearly Keplerian rotation
curve \citep{adams89}.
However, both the ionized and the atomic gas in \sgraw\ have densities
which are much too small to support a gravitational instability,
with Toomre Q parameters \citep{binney87} of $\sim$1000 and $\sim$50,
respectively.
The molecular gas in the CND could be gravitationally unstable if
its mass is as large as concluded by \citet{christopher05} and
\citet{montero09}.
However, if the smaller mass derived from \fir\ dust emission
\citep{etxaluze11} is assumed, even the CND should not support gravitational
instabilities.
The mass density of the stellar distribution is large enough to be
gravitationally unstable if it is flattened into a disk, but the
distribution of the older stars, which constitute the bulk of the
stellar mass, is probably not highly flattened.
We conclude that most likely the ionized spiral is not a result of a
density wave caused by gravitational interactions within the disk.
But a wave might be induced by another perturbing force,
perhaps due to magnetic fields.

Alternatively, a perturbing force may not be necessary to organize
the gas in \sgraw\ into a spiral pattern.
The reason that a one-armed spiral is the main instability in a
potential dominated by a point mass is that orbits are approximately
elliptical, with one focus at the center, so that if gas orbits are
eccentric with orientations varying with radius they crowd along a
one-armed spiral.
In addition, the presence of distributed mass in the star cluster
modifies the potential in such a way as to cause the orbits to precess,
which could cause their orientations to vary with radius.

To investigate the possibility that gas orbits in the Galactic center
gravitational potential naturally set up a spiral pattern we ran a
simulation of orbits in a mass distribution like that in Eq. 1.
We started the simulation with orbits along aligned ellipses,
with all ellipses having one focus at \sgras\ and a distance between
the two foci varying as $a^{\gamma}$, with $\gamma = 0-1$,
and with the orbital plane uniformly populated with particles
representing the gas.
Each particle was allowed to orbit in the potential of a black hole
plus a power-law stellar mass distribution, with no interactions
between the particles.
With this potential, orbits are well approximated with ellipses that
precess in the retrograde direction.
That is, the time from apocenter to apocenter is less than the time
for a 360$^{\circ}$ motion.
For a mass distribution power law steeper than r$^{-1.5}$ inner orbits
precess faster than outer, causing the orbits to crowd along a leading
spiral, whereas for a shallower power law outer orbits precess faster,
causing a trailing spiral.
The starting point of the simulation for an r$^{-0.5}$ stellar density
distribution is shown in Figure 11a, and the distribution of particles
after 1.4\ee{5}\,yr is shown in Figure 11b.
The spiral persists for several times longer than it took to form since
it wraps on a precession time scale, which is several $10^5$~yr, rather
than the orbital time scale, which for the smallest orbits is several
$10^3$~yr.

The spiral in the simulation strongly resembles that which we observe,
in that it is quite open (not tightly wrapped) and approximately linear
or Archimedean.  In addition, the initial conditions of our simulation
seem plausible as a situation that could result from the infall of a
molecular cloud into the central region, and the relatively flat stellar
density distribution required to produce a trailing spiral is consistent
with recent observations \citep{buchholz09, do09}.
We also note that the differential precession that causes the orbit
crowding only occurs within $\sim$2~pc of the center, consistent with
the lack of a prominent spiral pattern in the molecular gas in the CND.
Our main concern about this model is that in the time required for an
orbit at 1.5\,pc to precess by 360\degree , gas at 0.1\,pc will have
orbited around the center roughly 100 times.
Whether gas could orbit this many times without being disrupted by
stellar winds or other infalling clouds is unclear.

We also compared the spiral wave model to the kinematic distribution
observed with the \neii\ line by making a synthetic data cube with
Doppler shifts calculated for orbits along a set of ellipses with
orientations varying linearly with radius and eccentricities varying
with radius to a power.
The resulting orbits agree well with those in our simulation.
The orbital velocities were calculated assuming constant angular
momentum and energy along each ellipse.
For our calculations, we used the effective potential:
\begin{equation}
U=-GM_{\bullet}/r+\alpha r^\beta+l^2/(2r^2)
\end{equation}
which corresponds to a stellar mass distribution varying as $r^{\beta - 2}$.
With this synthetic data cube we could calculate the predicted Doppler
shift at each point in the sky, allowing us to spectrally shift and
collapse our observed data cube as we did for our circular orbit model,
and to search for parameters that give the best agreement between the
model and the data.
The relevant parameters are the orientation for $a=0$ ($\psi_{i}$), 
the rate of change of orientation (${d\psi}/{da}$),
and the rate of change in eccentricity
(given by $e_{1}$, the eccentricity at $a=1$\,pc, and the power, $b$, such that
$e=e_{1} (a/1\,pc)^b$), as well as the potential energy parameters,
$M_{\bullet}$, $\alpha$, and $\beta$.
We held the disk parameters constant at $\imath=66^{\circ}$
and $\Omega=15^{\circ}$.
The best fitting parameters were $\psi_{i}=118^{\circ}$,
$d\psi/da=214^{\circ}\,pc^{-1}$,
$e_{1}=0.3$, $b=-0.5$, $\alpha=500$, $\beta=1$,
and $M_{\bullet}=4.5\ee{6} \msun$.
This corresponds to a stellar mass within r = 1\,pc of 1.16\ee{5}\,\msun.
The spatial distribution of ellipses is shown plotted on the \neii\ data in
Figure 12a, and a P-V diagram extracted from along the
northern arm and the western arc is shown in Figure 12b.
Note that the density wave spiral model lies significantly outside the
\neii\ spiral, which causes the model P-V diagram in
the northern arm to turn to the red.
By adjusting the starting position, $\psi_{i}$, we can bring the model closer 
spatially to the data. 
With $\psi_{i}=175^{\circ}$, the density wave model lines up with the
\neii\ emission, and, perhaps as expected,
the P-V diagram looks like Figure 7a for motion along the spiral.
With $\psi_{i}=155^{\circ}$, the density wave peak 
lies just outside the \neii\ spiral (Figure 13), and the model
P-V pattern agrees well with the observations.
We choose to manually adjust this parameter to specifically fit the observed
spatial pattern.
The original fit ($\psi_{i}=118^{\circ}$) must have been affected by emission
that is not part of the spiral pattern, resulting in a worse fit for the P-V
diagram.
By adjusting $\psi_{i}$, we see a far superior fit to the spiral,
while presumably causing a worse fit for the material elsewhere in the map.
That the ionized gas peak would lie inside of the density peak (Figure 13) is sensible
if the gas becomes ionized on entering the spiral or if ionizing radiation
illuminates the inner edge of the spiral.
It is important to note that we now get a mass distribution similar to that
determined by stellar observers.
This is a result of the fact that the orbit crowding is strongest
just beyond apocenter where the particles are moving more slowly
than the circular speeds and are moving slightly inward, 
which is consistent with the slight inward velocity derived for the 
circular orbit model.

To test this model more thoroughly would require a hydrodynamic, or
perhaps a magneto-hydrodynamic simulation, which will have to be a
project for the future.

\subsection{\it What we don't explain}

One problem with a density wave model of the ionized spiral
is explaining the high contrast seen, especially between the northern
arm and the region just to its west, where the emission is a factor
$\sim$100 fainter.  A possible explanation is that the gas is not
just compressed in the spiral, but is also more highly ionized there.
The gas could be largely neutral over most of its orbit, but become ionized
as a result of passing through a shock or by interacting with the
magnetic field when entering the northern arm and western arc.
Although stellar ultraviolet radiation appears to be sufficient to
account for the ionization of the gas in \sgraw, the fact that the
stars can only be observed in the infrared, making it difficult to
determine their ultraviolet luminosities, leaves open the possibility
of contribution by other sources of ionization.
We can place an upper limit on the rate of conversion of kinetic energy
into ionization in a shock by estimating the rate at which gas
carries kinetic energy into the northern arm.
If we estimate the mass of (predominantly neutral) gas in the inner
1.5 pc to be 1000 \msun, take a typical speed of 100 \kms, and assume
1/4 of its kinetic energy is dissipated once per orbit, or each time
it passes through the spiral pattern, we estimate a power of 2\ee{4} \lsun.
This is much less than the ionizing luminosity of $\sim 10^6$\lsun\
needed to maintain the ionization \citep{brown84}.
Apparently shocks make at most a minor contribution to the ionization
of the gas.

We also have not explained the orientation of the magnetic field that
runs along the northern arm.
We (perhaps naively) would have expected the field lines to be aligned
with the gas motions, which we conclude run diagonal to the northern
arm.  But compression of a field in a spiral shock may instead cause
the field to be aligned parallel to the shock.
A magneto-hydrodynamic simulation may be able to resolve this question.

Finally, we note that we have not proposed a model of the eastern arm and
bar region, which account for about half of the ionic emission from \sgraw.
The elliptical orbit model of \citet{zhao09} fits the spatial distribution
of these features rather well, but does not agree well with the observed
\neii\ kinematics.
Neither are these features fitted by circular motions of gas in a plane.
As a speculative suggestion, we note the morphological similarity of the
eastern arm, especially its northeastern loop (which is most prominent
in the VLA free-free maps) to solar prominences.
Perhaps magnetic fields very close to the center are strong enough to
lift ionized gas out of the center along this feature.
A measurement of the magnetic field in this region would be difficult
due to the faintness of the dust emission there, but would be of interest.

\section {SUMMARY}

The \neii\ observations strongly favor a model in which much of the
ionized gas in \sgraw\ orbits in a plane close to that of the CND,
with nearly circular motions.
The spatial pattern of the gas in this plane is well described by
an approximately linear spiral, which includes the western arc and
the northern arm.
This model requires the gas orbits to cross the spiral, especially in
the northern arm region.

We have considered several implications of and physical models to
explain these conclusions.
The most promising physical model involves a spiral density wave
resulting from the precession of elliptical orbits in the potential
of the central super-massive black hole and star cluster.
In addition to providing an explanation for the origin of the spiral,
this model results in a best-fitting mass distribution in agreement
with that derived from orbital motions of stars around the SMBH,
and with relatively flat stellar mass distribution in the inner parsec.

\acknowledgments

We thank Milos Milosavljevic for very helpful discussions
and Thomas Greathouse for assistance with the observations.
We also thank Mar\'{i}a Montero-Casta\~{n}o for providing the
HCN data.
This work was supported by NSF grants AST-0607312 and AST-0708074.

\clearpage

\begin{figure}
\epsscale{0.8}
\plotone{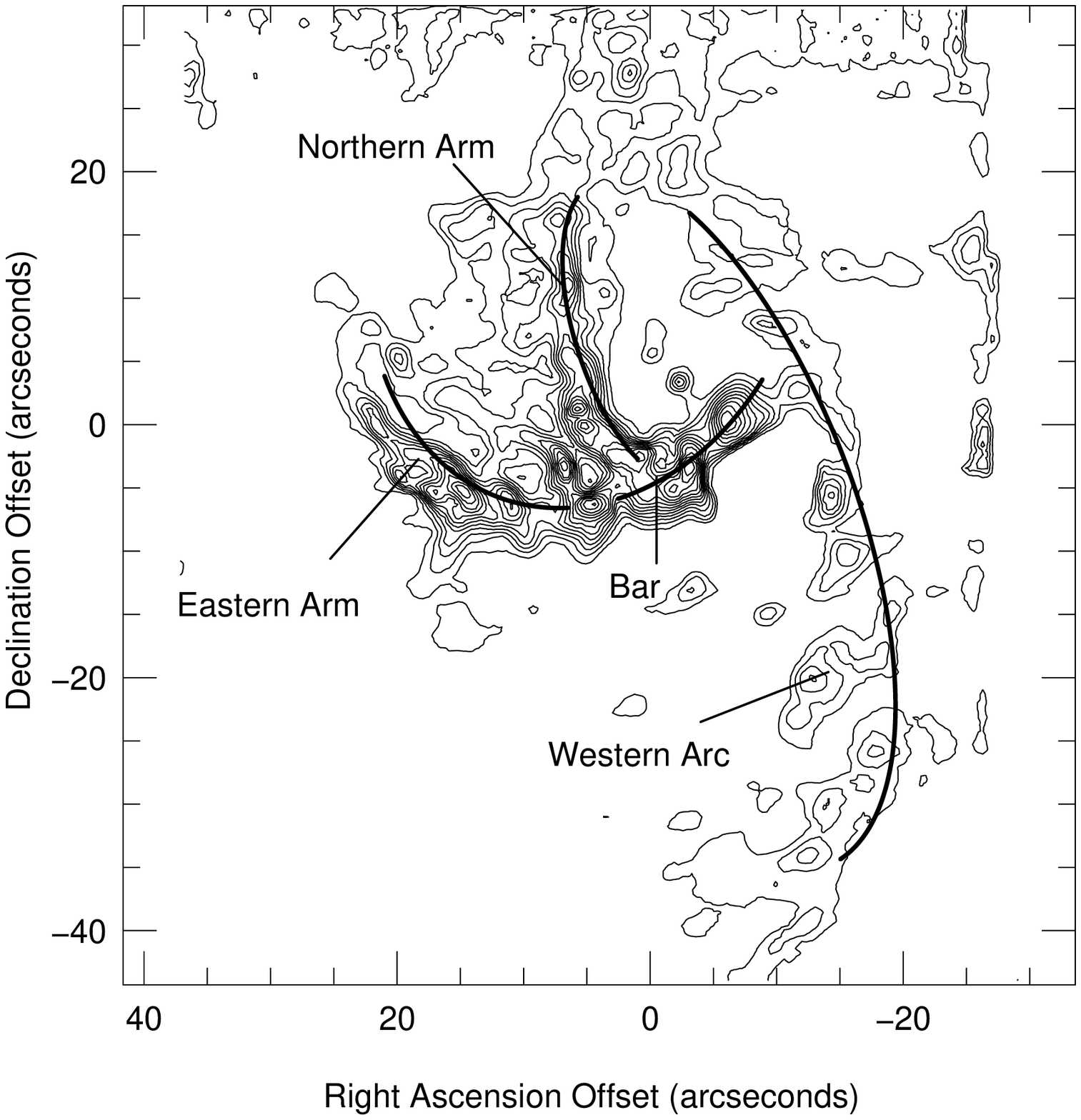}
\caption{Contour plot of \neii\ emission, with a nonlinear stretch, shown here
with the various structures labeled.
Offsets are from \sgras\ at 17$^h$45$^m$40.04$^s$
-29\degree 00\amin 28.11\asec\ (J2000).
\newline
(The \neii\ data cube is available as a FITS file in the on-line journal.)}
\end{figure}

\begin{figure}
\epsscale{0.8}
\plotone{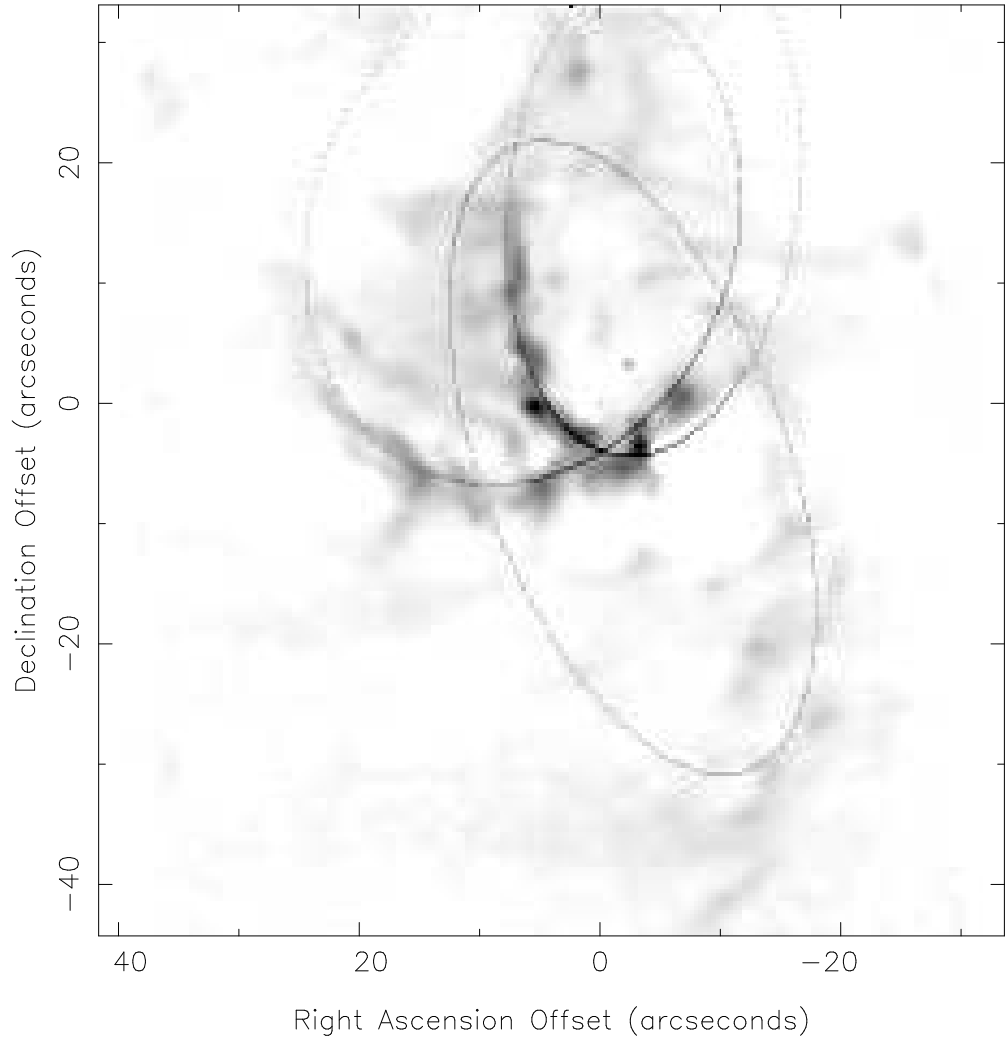}
\caption{Integrated \neii\ emission with a nonlinear stretch and the
elliptical orbits of \citet{zhao09} superposed.}
\end{figure}

\clearpage

\begin{figure}
\begin{center}
\hspace{0.12in}
\includegraphics[angle=270, scale=0.5]{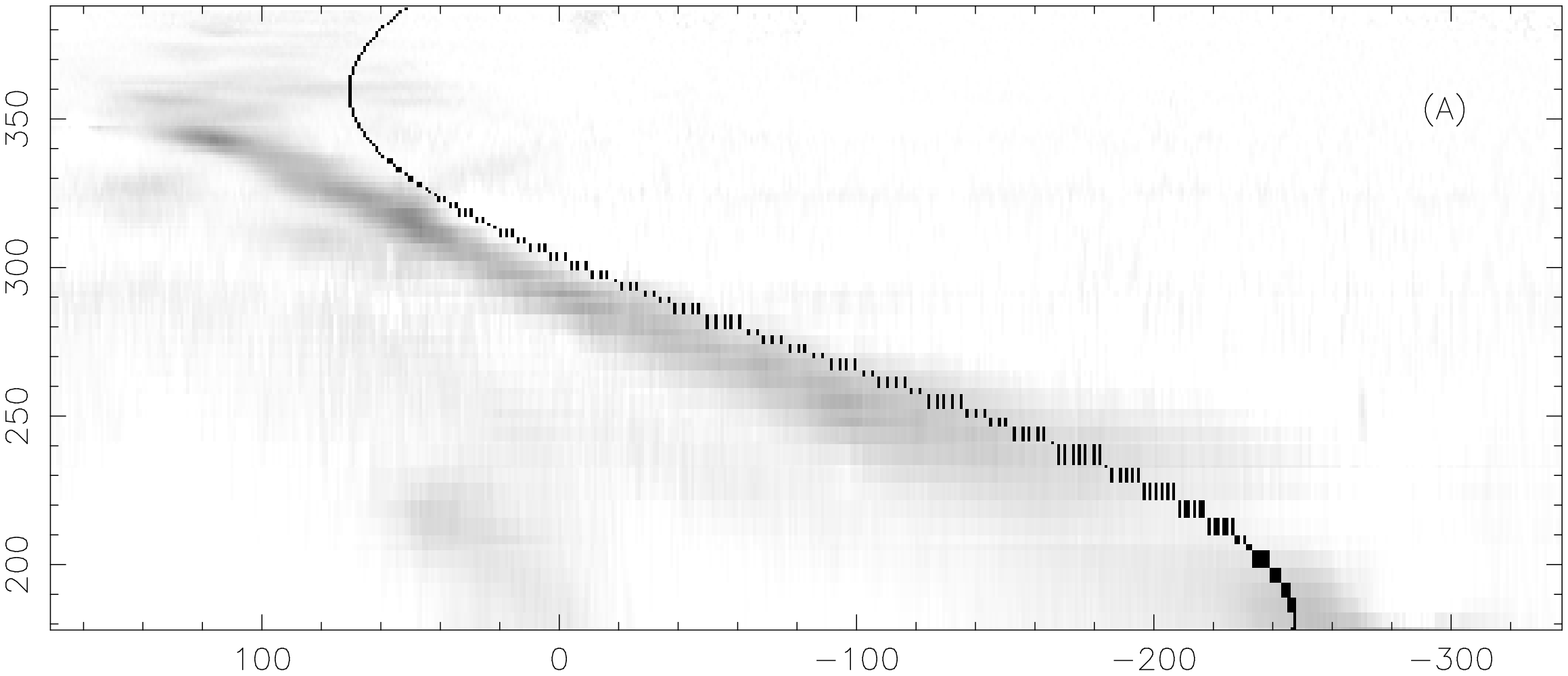}\\
\vspace{0.2in}
\includegraphics[angle=270, scale=0.5]{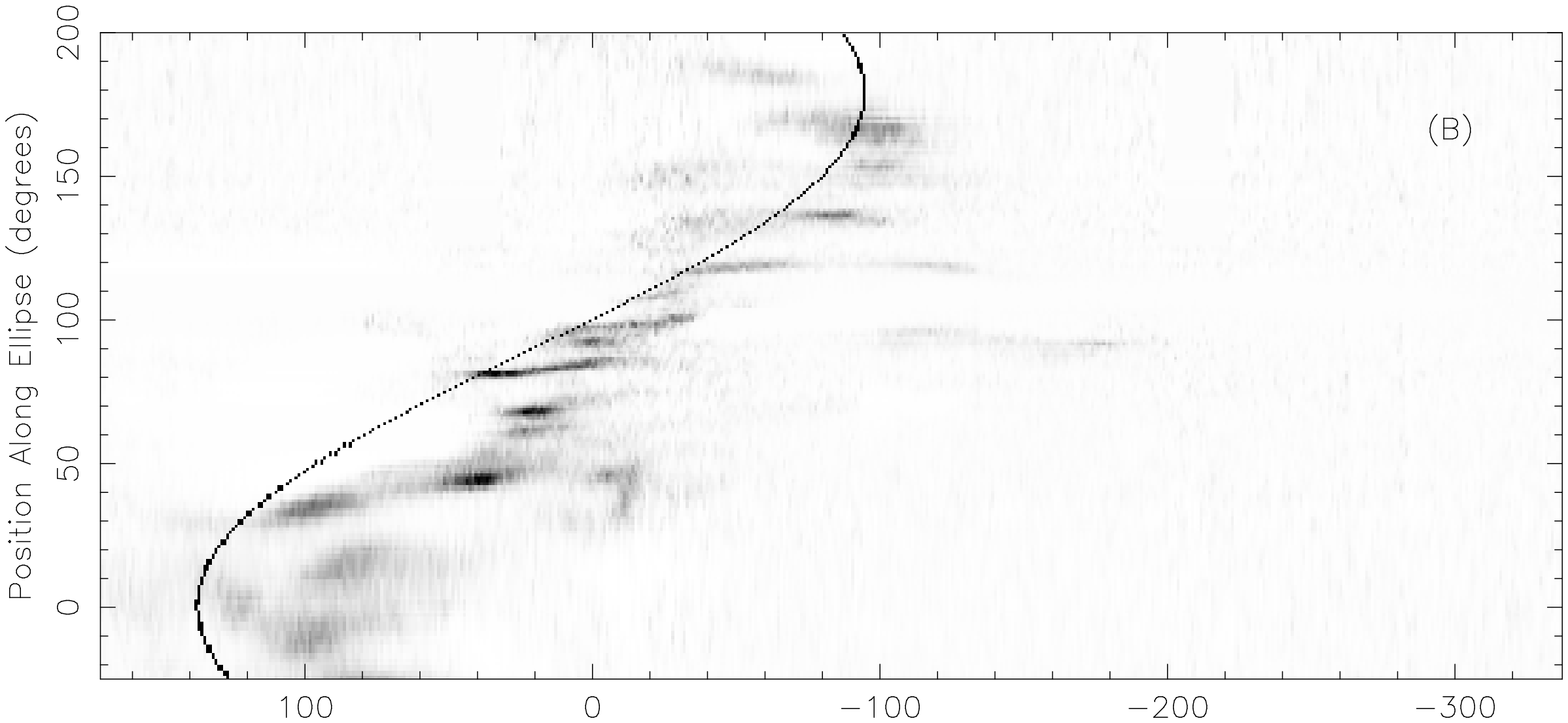}\\
\vspace{0.25in}
\hspace{0.12in}
\includegraphics[angle=270, scale=0.5]{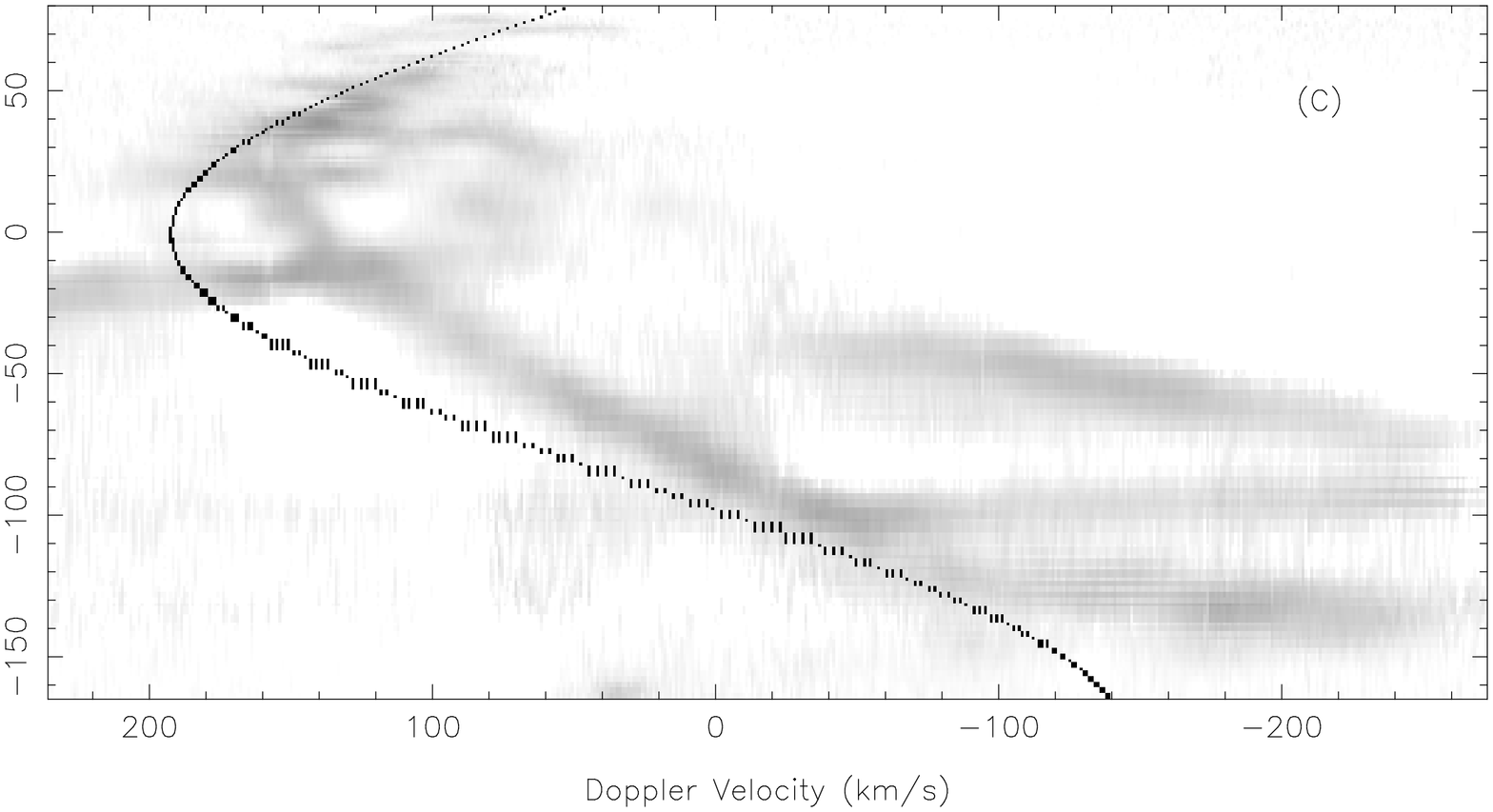}
\caption{Position-velocity diagrams extracted from the \neii\ data cube along the ellipses, 
along with the calculated velocity patterns for the ellipse model.  
Vertical axes are linear in angle around the ellipses running clockwise along the streamers.
Top to bottom: the northern arm, the western arc, and the eastern arm.}
\end{center}
\end{figure}

\clearpage

\begin{figure}
\epsscale{0.8}
\plotone{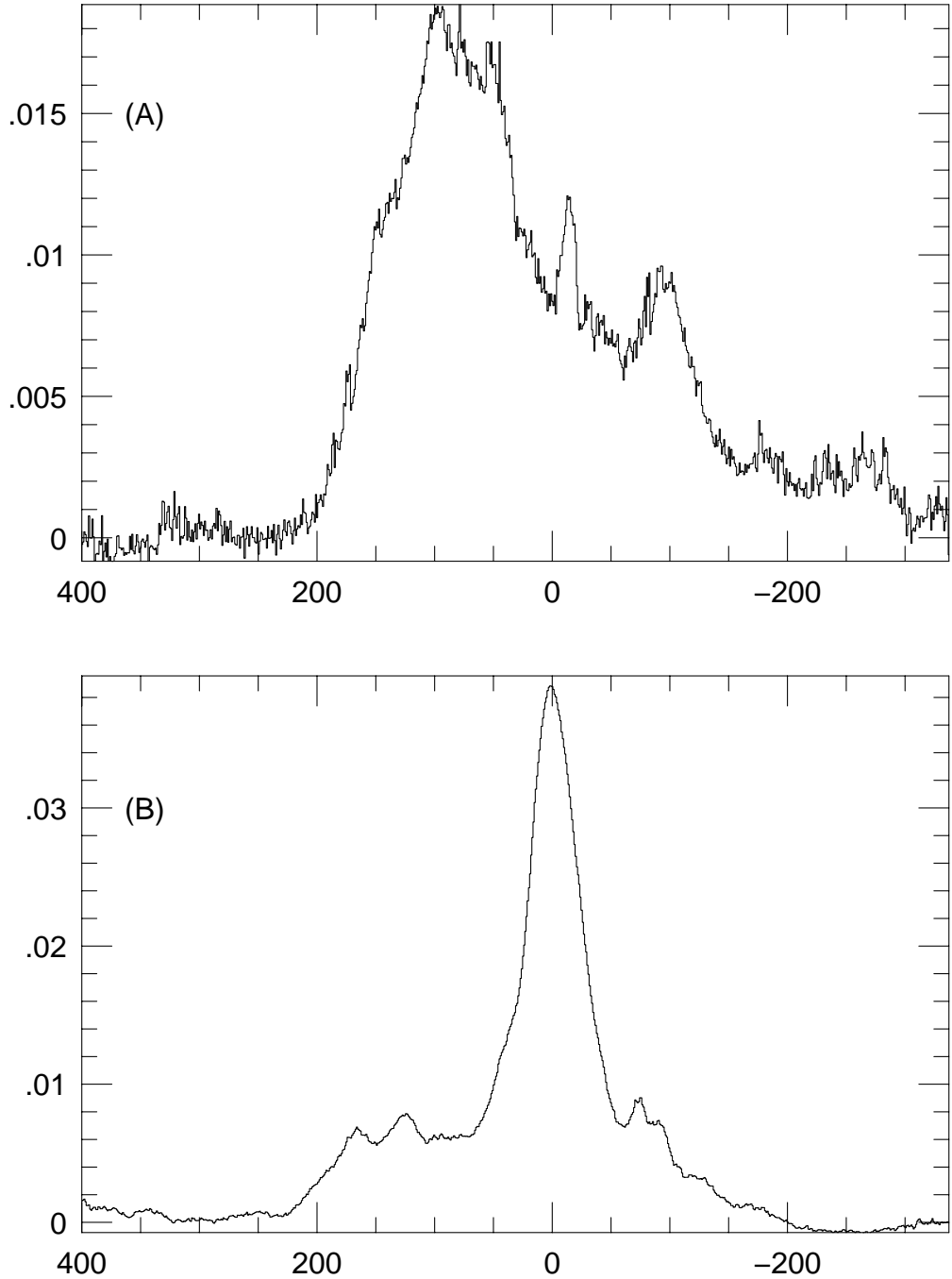}
\caption{a) Spectrum taken by collapsing the data cube spatially before shifting.
Here, 15\% of the \neii\ emission fits within 10\% of the velocity range
centered at zero Doppler velocity.
b) Spectrum after shifting the data cube to fit circular motion.
Now, 45\% of the \neii\ emission fits within 10\% of the velocity range
centered at zero Doppler velocity.
This is the "goodness of fit" parameter for the routine.
The horizontal axis is Doppler velocity (\kms).
The smoother appearance of 4b is a result of shifting and averaging
systematic noise from the background subtraction.}
\end{figure}

\begin{figure}
\plotone{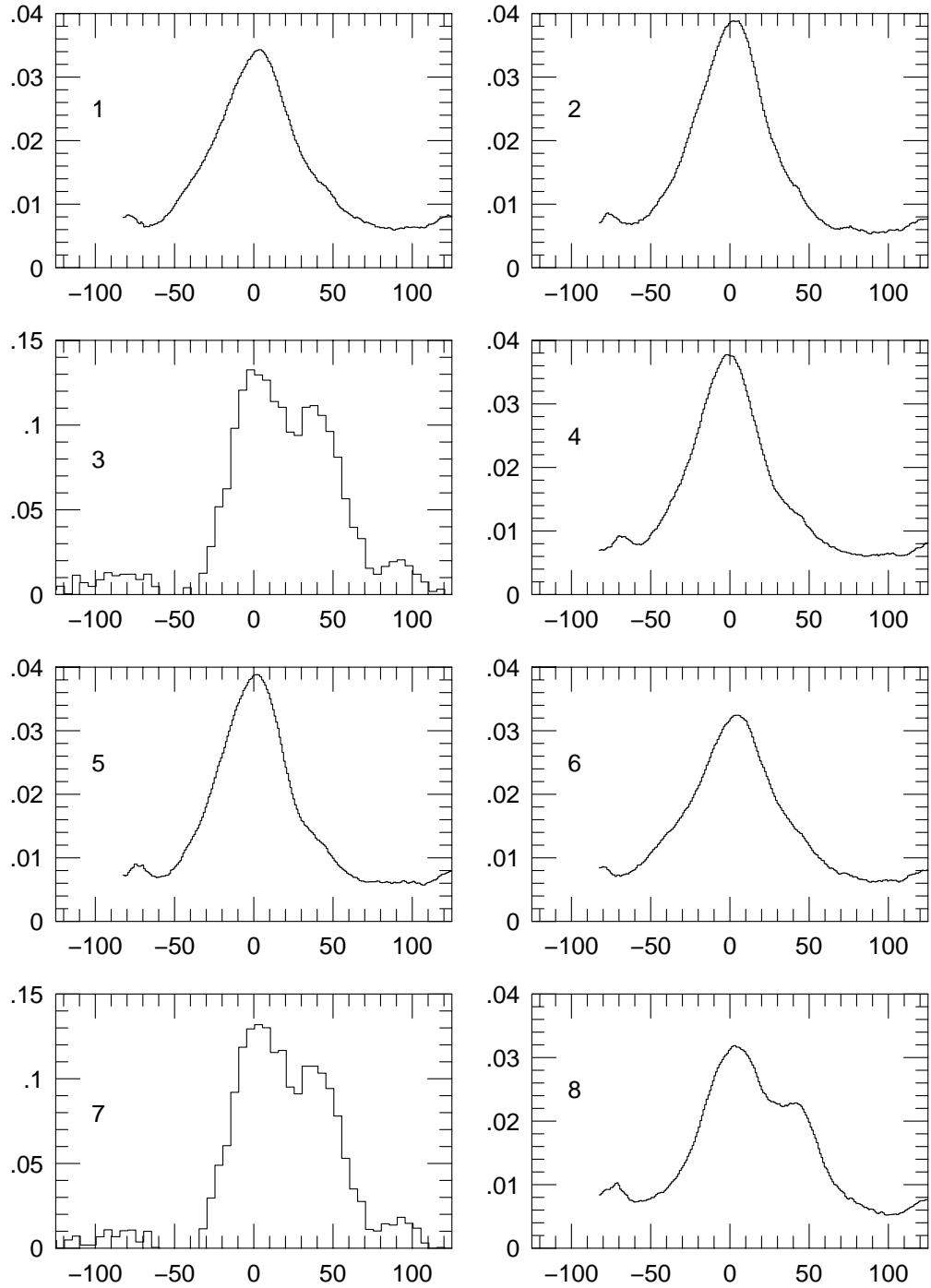}
\caption{The shifted and collapsed spectra corresponding to the various best
fits in Table 1. The horizontal axes are Doppler velocities in \kms.
Panels 3 and 7 are for the \citet{montero09} HCN data.}
\end{figure}

\clearpage

\begin{figure}
\begin{center}
\includegraphics[scale=0.45]{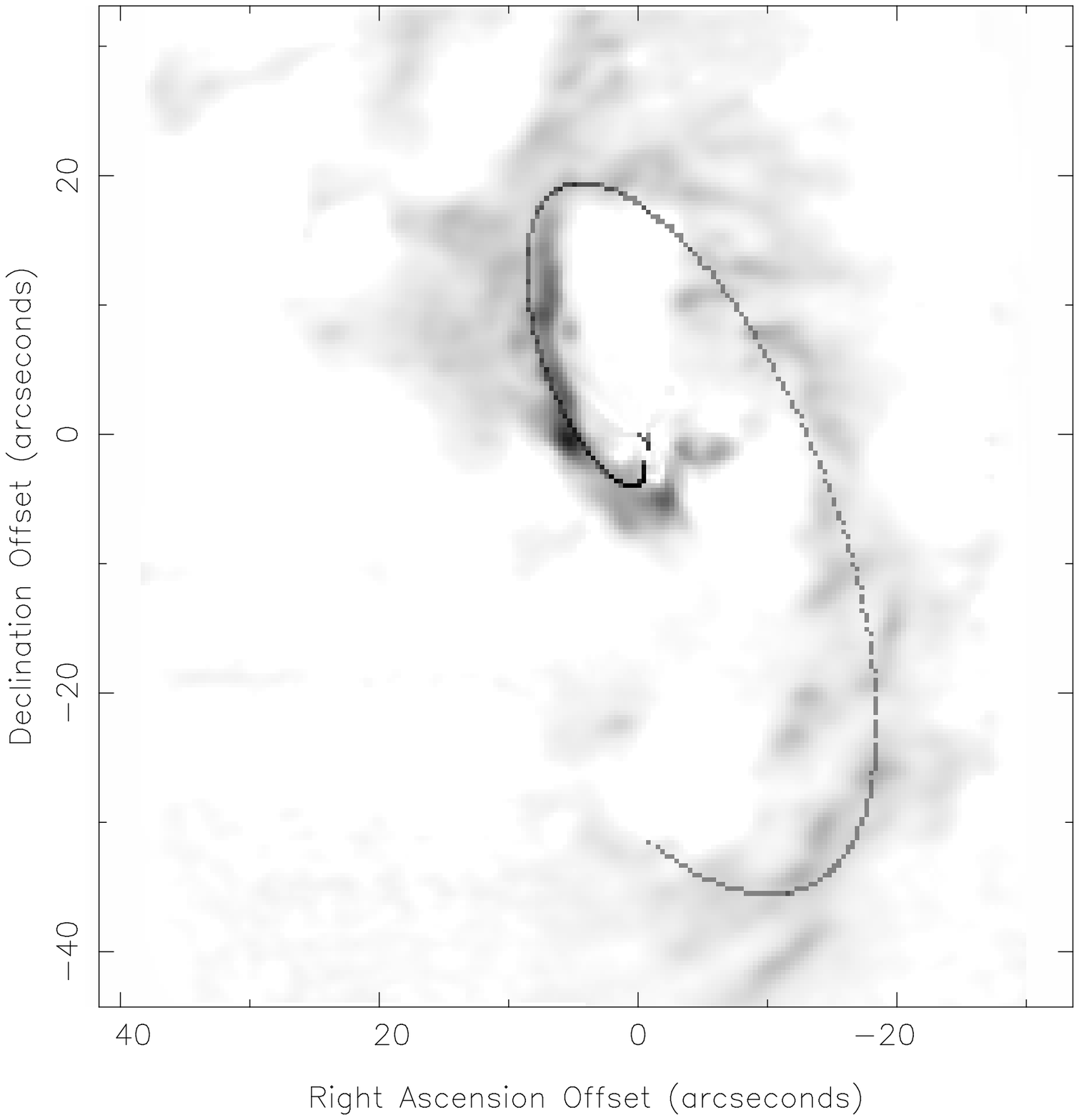}
\includegraphics[scale=0.45]{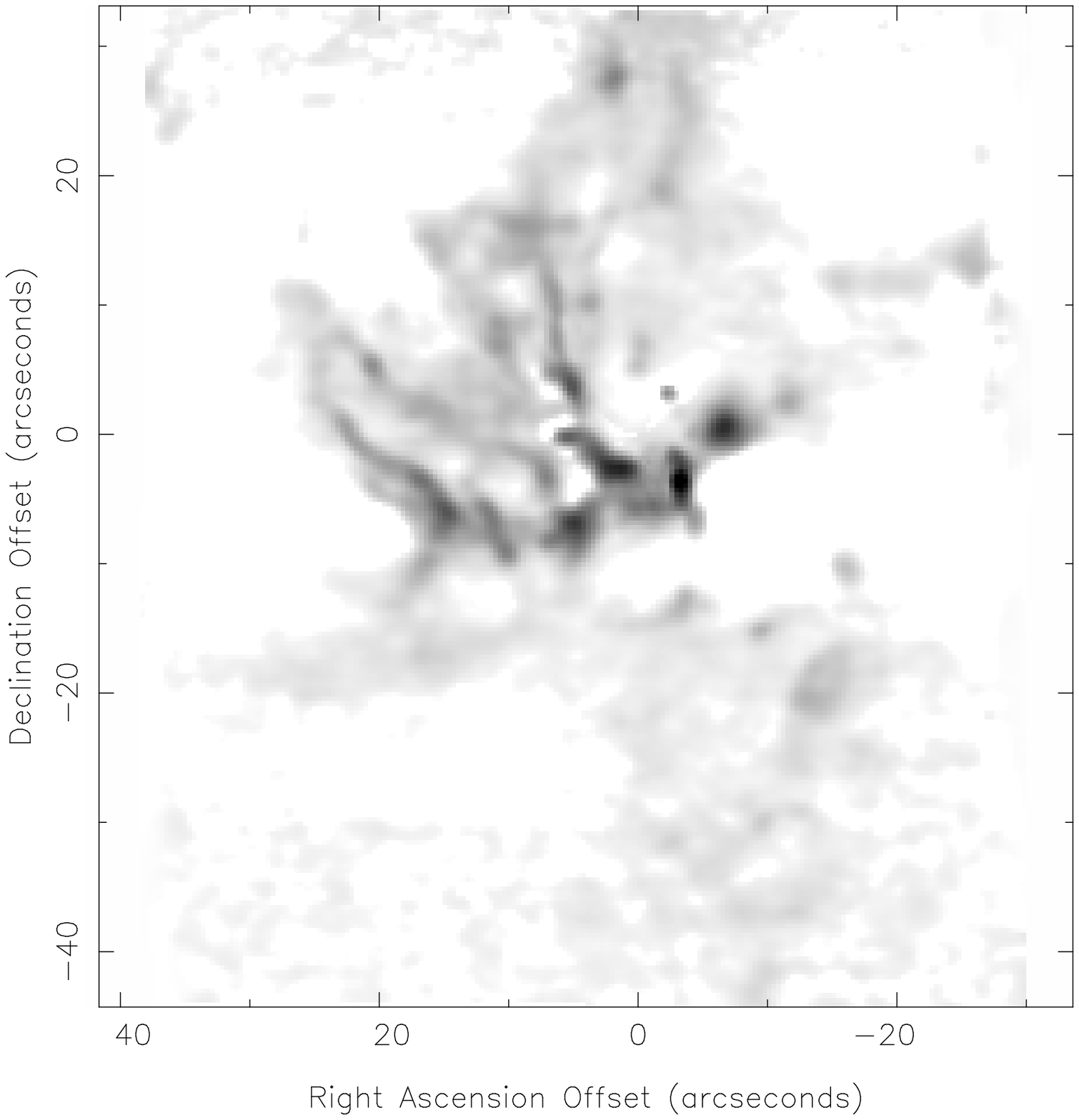}
\caption{Left: Grayscale image of the ionized gas that fits a nearly
circular orbit model
in the plane defined by $\imath = 66^{\circ}$, $\Omega = 23^{\circ}$, with 
$a = -0.06$\,pc, within $\pm$30\,\kms.
The spiral model is superposed.
Right: The emission in the \neii\ data cube that does not
fit the circular orbit model.  That is, emission within $\pm$30\,\kms\ has been
masked out.  Both figures are shown with a square root stretch.}
\end{center}
\end{figure}

\clearpage

\begin{figure}
\begin{center}
\includegraphics[scale=0.5, angle=270]{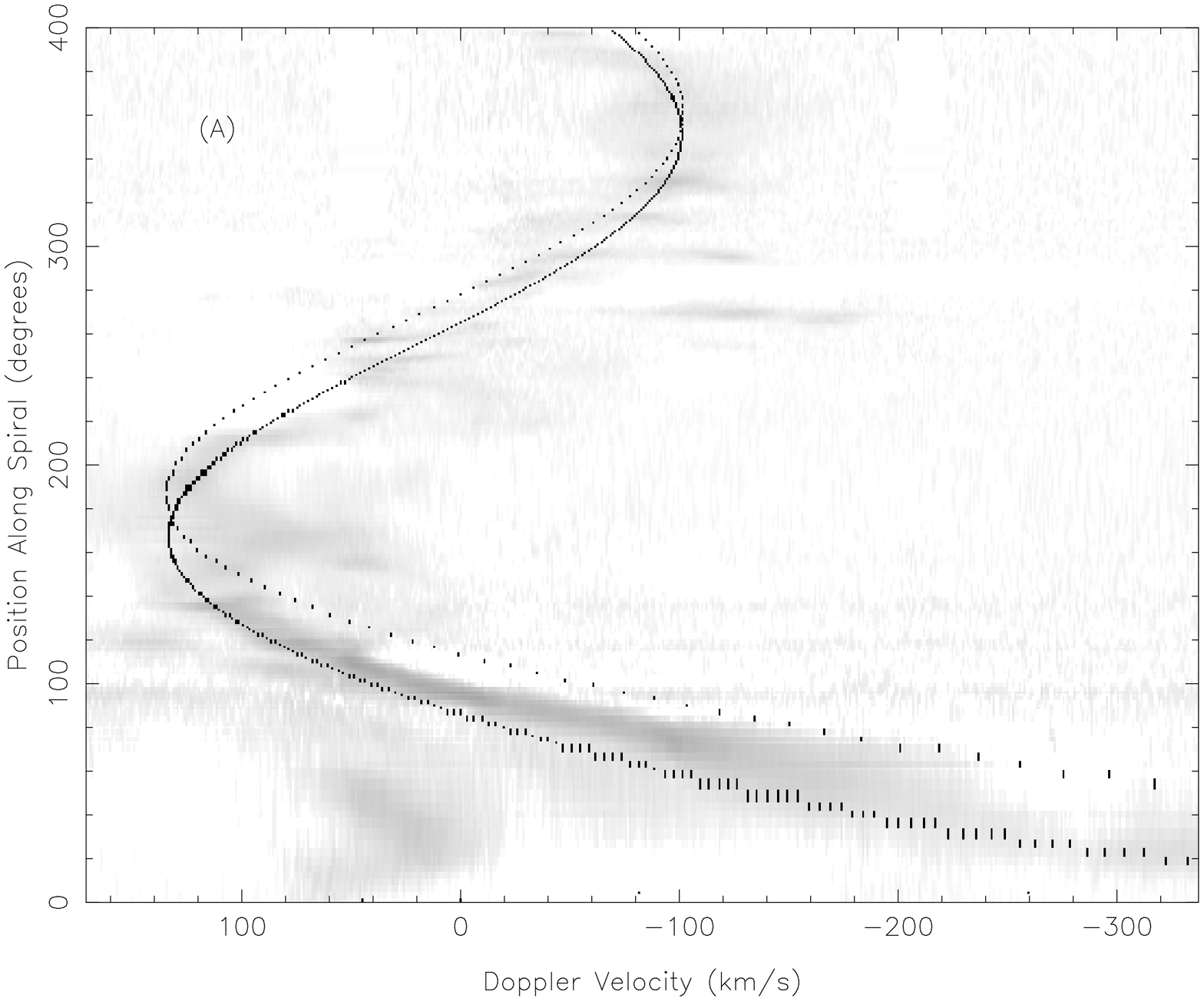}
\includegraphics[scale=0.5, angle=270]{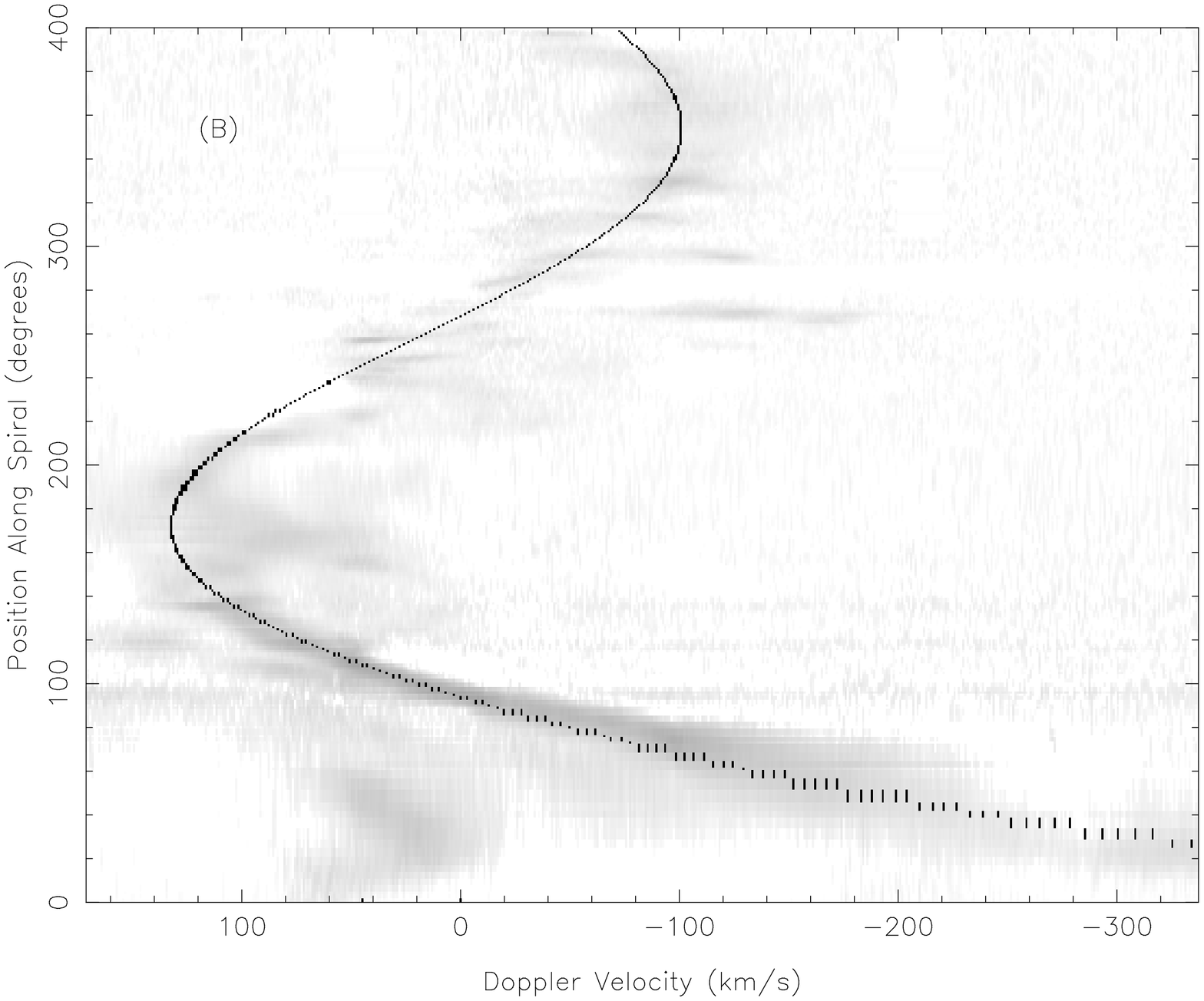}
\caption{a) Position-velocity diagram extracted from the \neii\ data cube
along the spiral with calculated velocity patterns for purely circular motion,
(using parameters from row 5 in Table 1 but with $a=0$\,pc, tagged every
1$^{\circ}$) and for motion along the spiral (using parameters from row 5
in Table 1 but with $a=-0.27$\,pc, tagged every 3$^{\circ}$) superposed.
b) Position-velocity diagram extracted from the \neii\ data cube along
the spiral with calculated velocity patterns for the best fit (using parameters from 
row 5 in Table 1).}
\end{center}
\end{figure}

\clearpage

\begin{figure}
\plotone{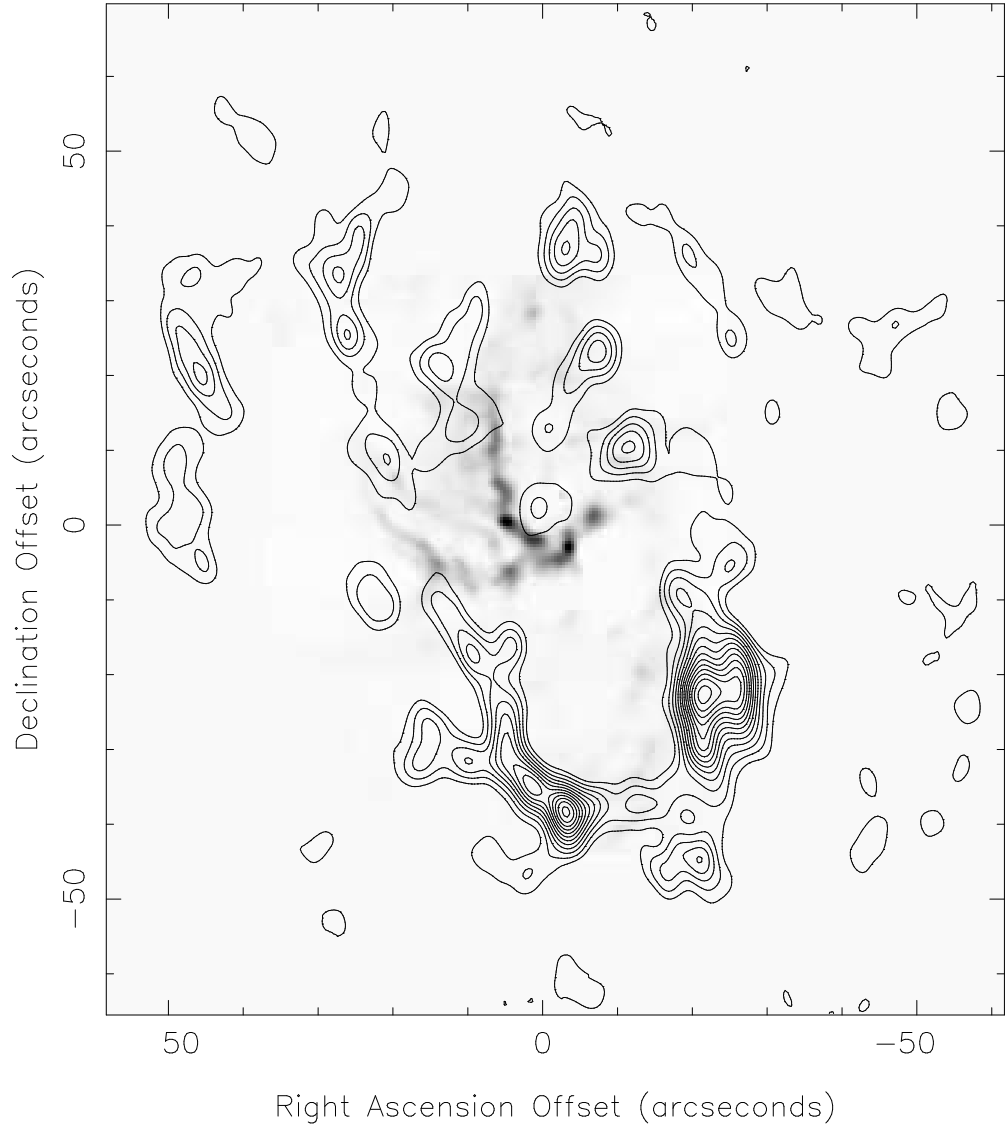}
\caption{Grayscale image of \neii\ emission summed over Doppler shifts from
$-339$\,\kms\ to  $+299$\,\kms\ using a nonlinear stretch, superposed on a
contour map of HCN(4-3) emission from \citet{montero09}
summed from $-90$\,\kms\ to $+130$\,\kms.}
\end{figure}

\clearpage

\begin{figure}
\epsscale{0.8}
\plotone{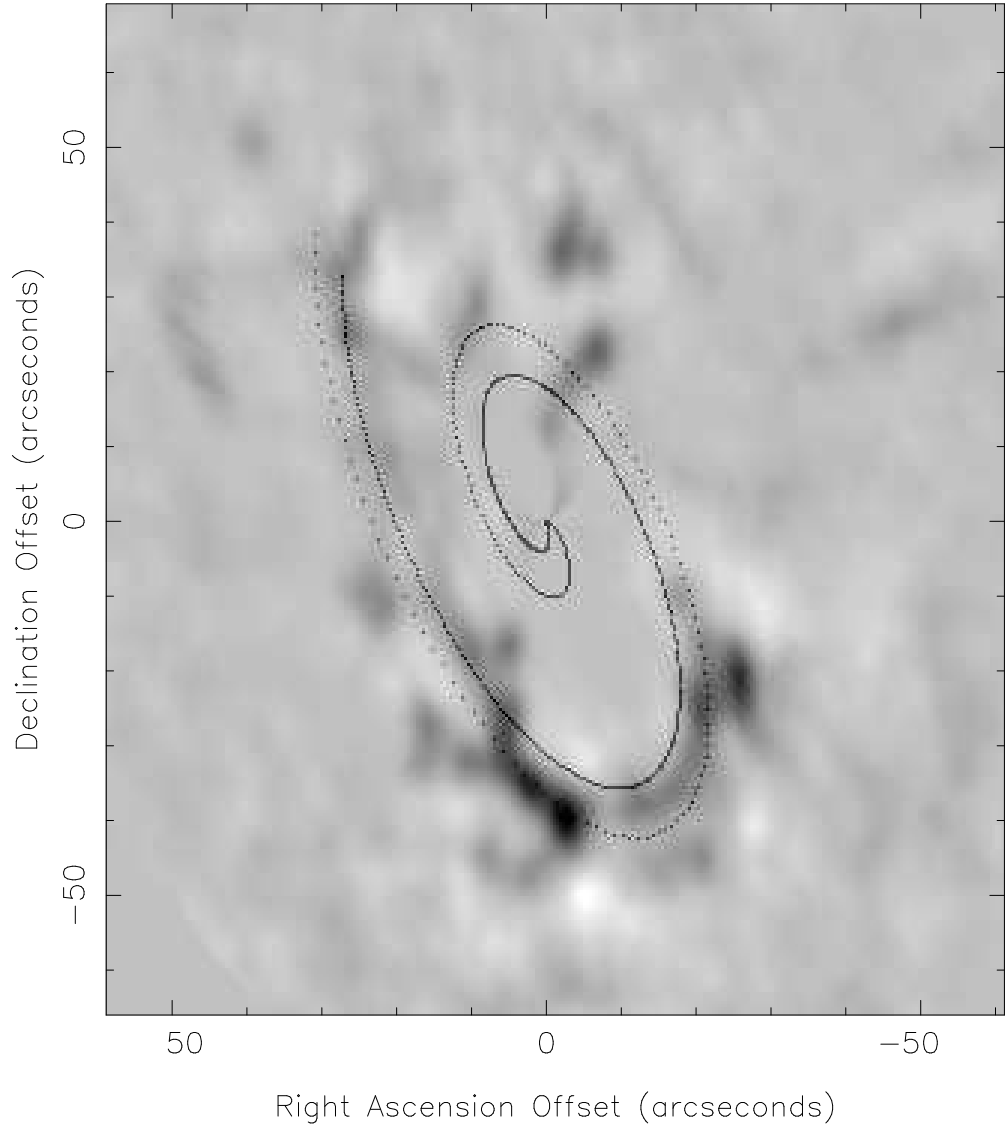}
\caption{HCN(4-3) emission map showing only the emission
that fits circular velocities within $\pm30$\,\kms\ in the plane described
by $\imath=66^{\circ}$ and $\Omega=23^{\circ}$, 
with the original spiral (1$^{\circ}$ steps, $\phi = 274^{\circ}$)
and a slightly larger spiral (3$^{\circ}$ steps, $\phi = 199^{\circ}$)
superposed.}
\end{figure}

\begin{figure}
\includegraphics[angle=270]{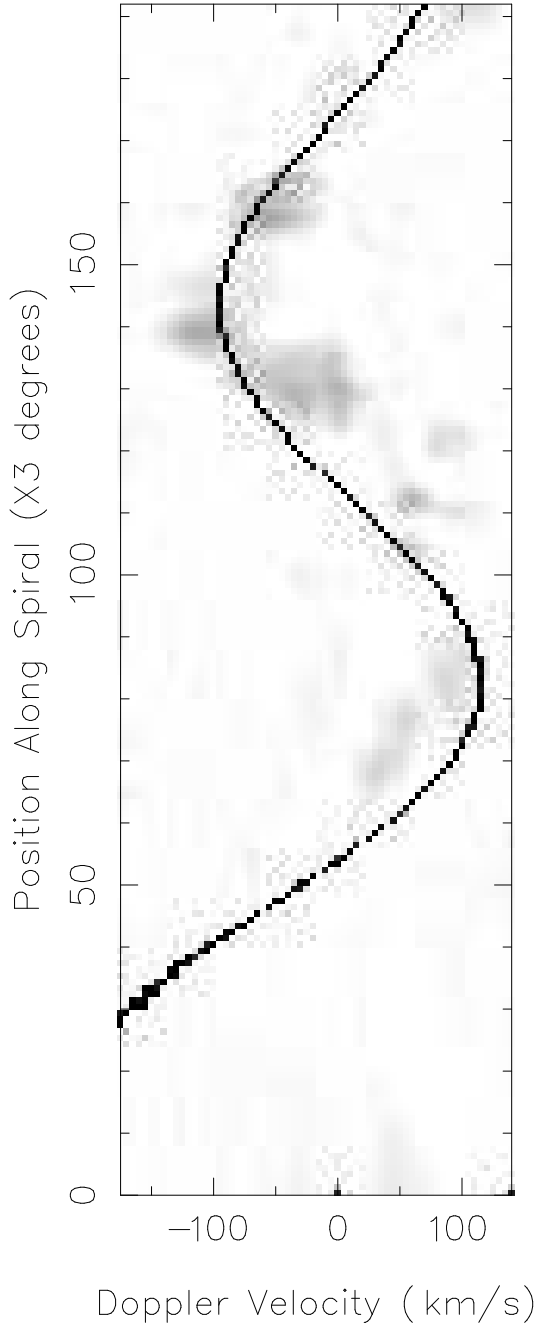}
\caption{Position-velocity diagram of HCN(4-3) emission along the outside
spiral in Figure 10 ($\phi = 199^{\circ}$);
the vertical axis runs from 0$^{\circ}$ to 576$^{\circ}$ in 3$^{\circ}$
steps; the horizontal axis runs from $-175\,\kms$~ to $+140\,\kms$.}
\end{figure}

\clearpage

\begin{figure}
\includegraphics[angle=270,scale=1.0]{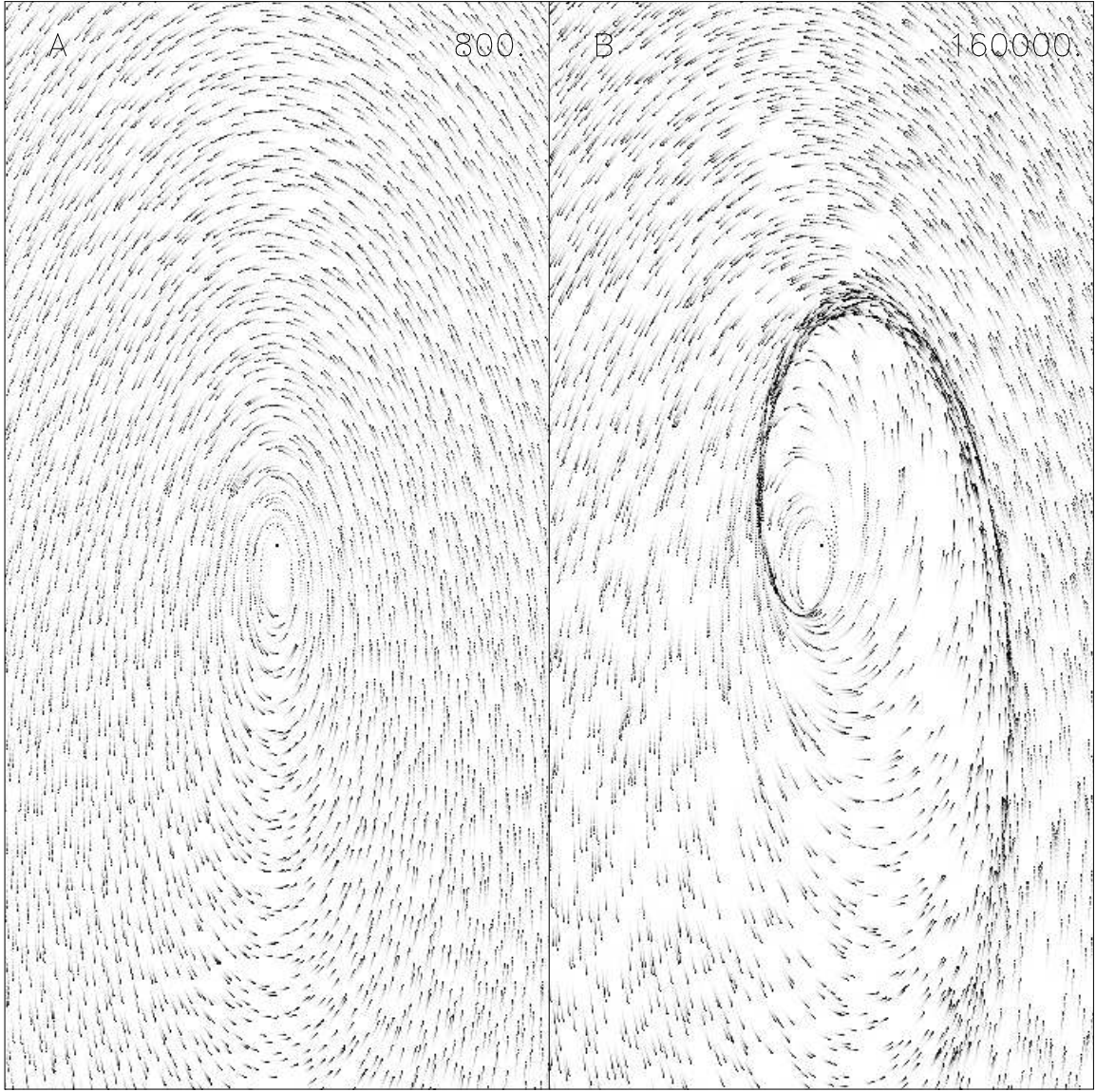}
\caption{a) Particle positions (with motions indicated with trails)
just after the start of a simulation of orbits in the potential of
a black hole and a flat stellar density distribution
with \eten{5}\msun\ contained within 1 pc.
The orbital plane is inclined by 65\degree\ about a vertical axis.
b) Particle positions after 1.6\ee{5} yr.
The spiral wave is a result of the differential precession of the
orbits; no interactions between the particles were included.}
\end{figure}

\clearpage

\begin{figure}
\begin{center}
\includegraphics[scale=0.55]{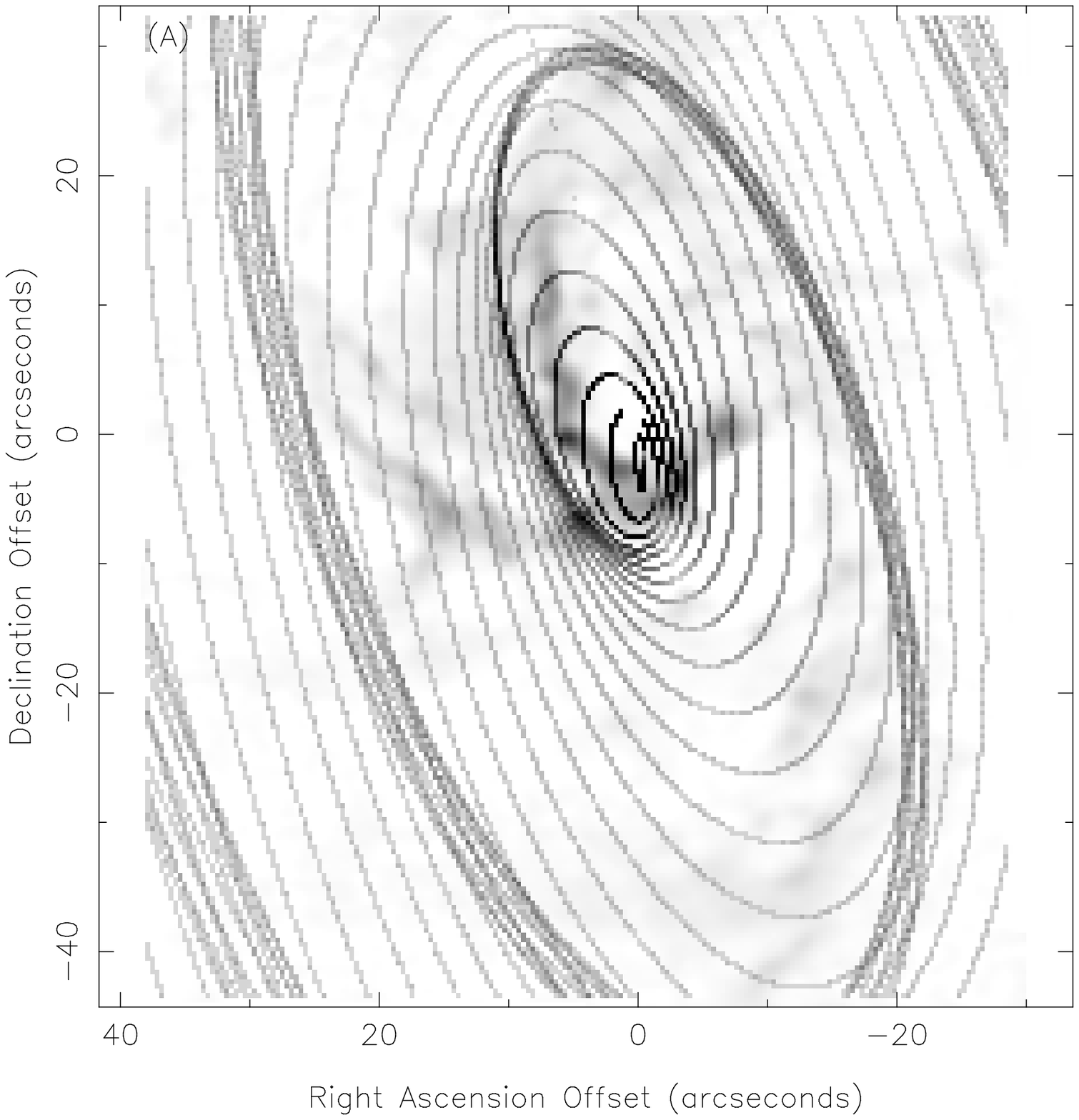}
\includegraphics[angle=270,scale=0.45]{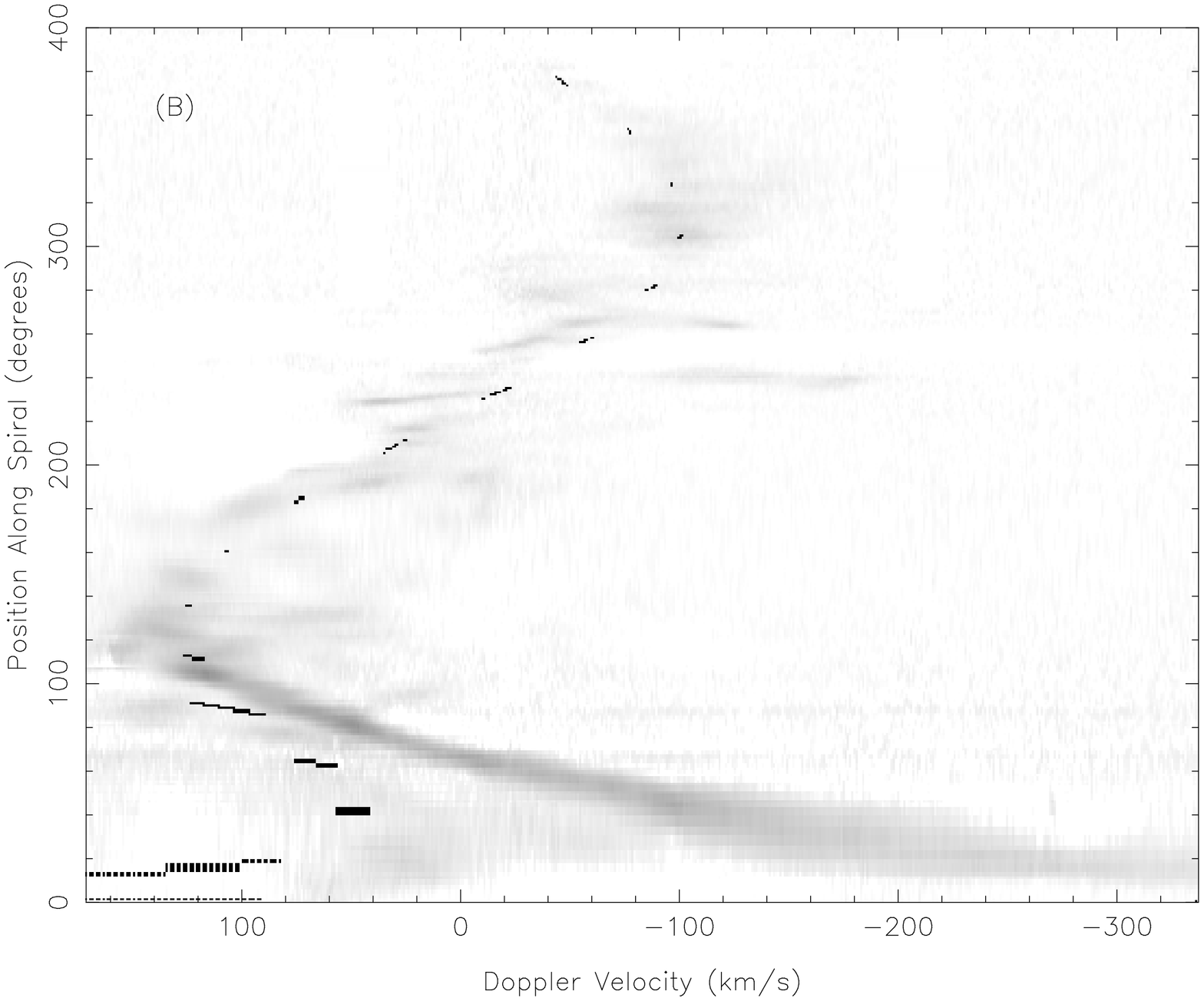}
\caption{a) The density wave model which gives the best fit to velocities
in the entire data cube.
b) Position-velocity diagram extracted from along the \neii\ spiral.
The parameters for this fit are:
$\psi_{i}=118^{\circ}$, $d\psi/da=214^{\circ}\,pc^{-1}$,
$e_{1}=0.3$, $b=-0.5$, $\alpha=500$, $\beta=1$,
and $M_{\bullet}=4.5\ee{6} \msun$.
The spatial fit is poor as the model spiral is far outside the \neii\ spiral. 
This effects the P-V diagram in that the \neii\ spiral crosses
to the other side of the innermost ellipses, causing the spectral model 
to turn to the red.}
\end{center}
\end{figure}

\clearpage

\begin{figure}
\begin{center}
\includegraphics[scale=0.55]{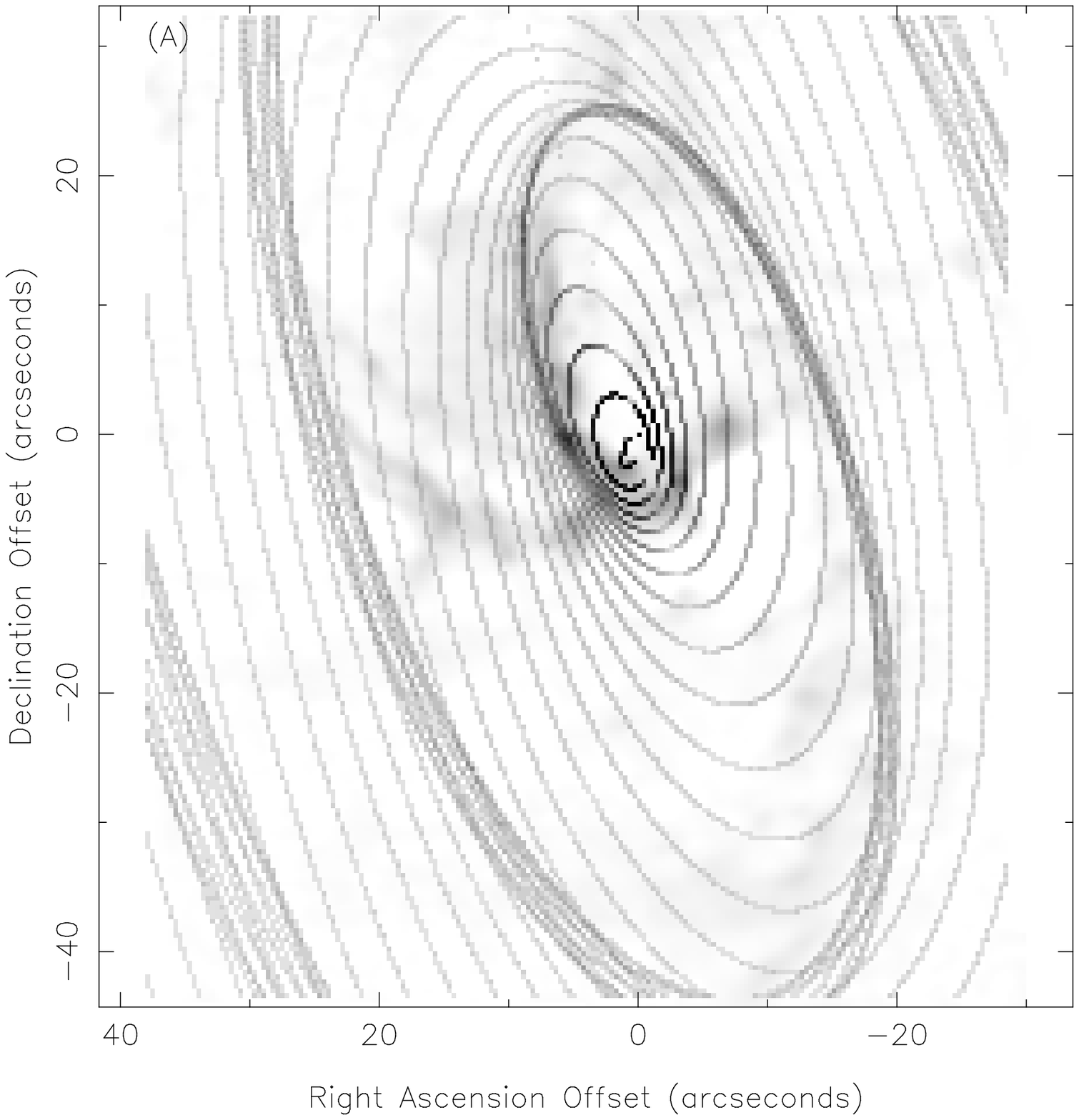}
\includegraphics[angle=270,scale=0.5]{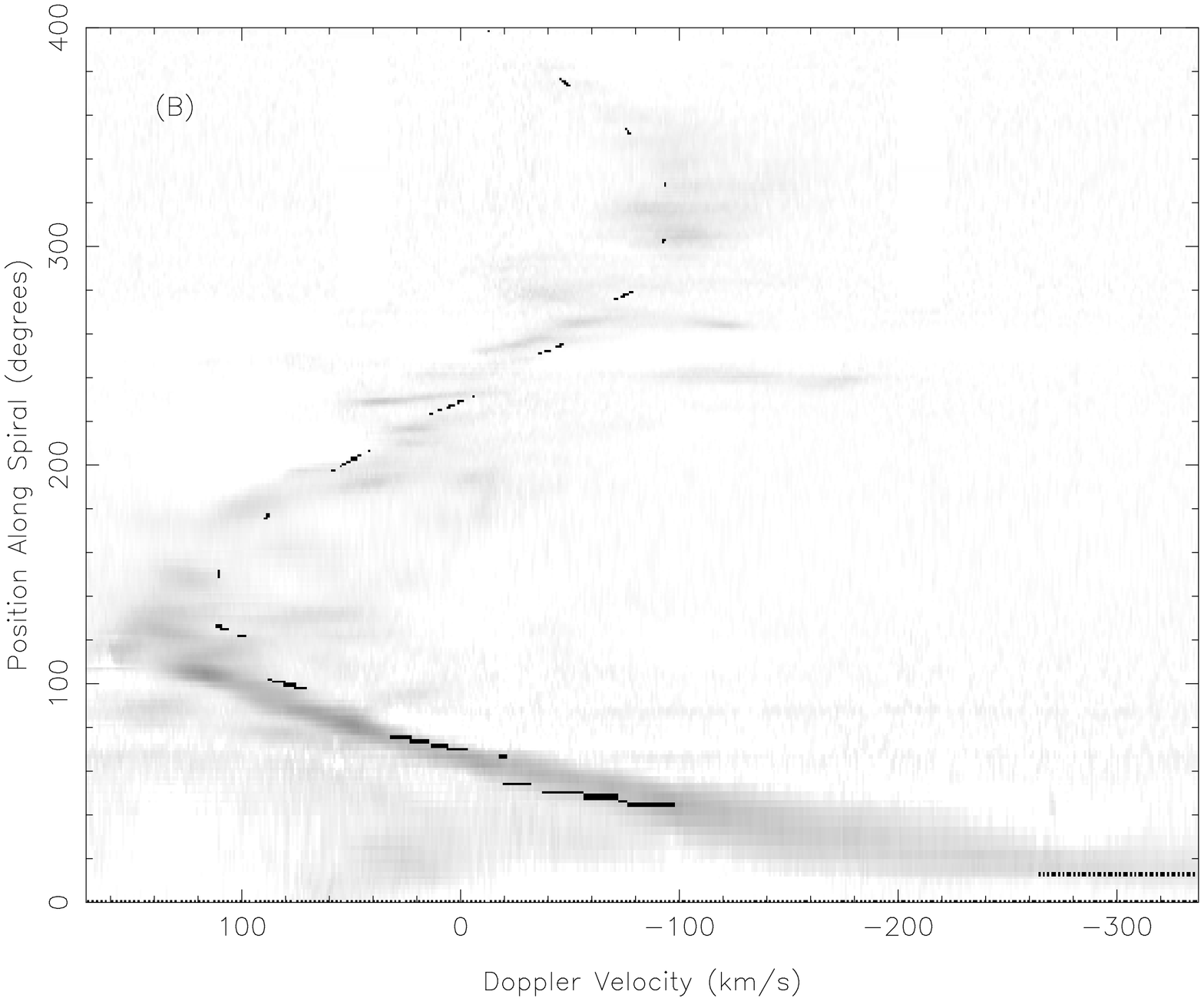}
\caption{a) The density wave model with the starting ellipse orientation
adjusted to achieve the best fit to the spectra along the spiral.
b) Position-velocity diagram extracted from along the ionized gas spiral.
Now, $\psi_{i}=155^{\circ}$.
This shows a very good spectral fit, and the spatial fit shows a density
peak just outside of the emission spiral.}
\end{center}
\end{figure}

\clearpage

\begin{deluxetable}{lcccccccc}
\tabletypesize{\scriptsize}
\tablecaption{Table of parameters}\label{tbl-1}
\tablewidth{0pt}
\tablehead{
\colhead{Stellar Distribution} & \colhead{$M_{\bullet}$ ($10^{6}$ \msun)} & 
\colhead{$\rho_{\circ}$ ($10^{6}$ \msun~$pc^{-3}$)} & \colhead{$R_{c}$ (pc)} & 
\colhead{$\imath$\tablenotemark{a}} &
\colhead{$\Omega$\tablenotemark{b}} & \colhead{$a$\tablenotemark{c} (pc)} &
\colhead{Goodness parameter\tablenotemark{d}}\\
\colhead{} & \colhead{} & \colhead{or $M_{1}$ ($10^{6}$\,\msun)} &
\colhead{or $\alpha$} & \colhead{} & \colhead{} & 
\colhead{} & \colhead{}}
\startdata
1) Eq. 1 ($M_{\bullet}$ fixed) &4.2 &1 &1.2  &70$^{\circ}$ &21$^{\circ}$ &-0.05 &0.41\\
2) Eq. 1 ($M_{\bullet}$ free) &2.6 &1 &1.2 &63$^{\circ}$ &24$^{\circ}$ &-0.07 &0.45\\
3) Eq. 1 (HCN data, $M_{\bullet}$ fixed) &4.2 &1 &1.2 &78$^{\circ}$ &29$^{\circ}$ &-0.42 &0.60\\
4) Eq. 3 ($M_{\bullet}$ fixed) &4.2 &0\tablenotemark{e}  &0  &68$^{\circ}$ &23$^{\circ}$ &-0.06 &
     0.43\\
5) Eq. 3 ($M_{\bullet}$ free) &3.5 &0.25 &0.5 &66$^{\circ}$ &23$^{\circ}$ &-0.06 &0.45\\
6) Eq. 3 (All Mass Fixed) &4.2 &1.0 &0.5 &72$^{\circ}$ &21$^{\circ}$ &-0.04 &0.39\\
7) Eq. 3 (HCN data, $M_{\bullet}$ fixed) &4.2 &0.25 &0.75 &77$^{\circ}$ &28$^{\circ}$ &-0.38 &0.61\\
8) Eq. 3 (a fixed)\tablenotemark{f} &1.4 &0.25 &2.75 &66$^{\circ}$ &42$^{\circ}$ &-0.27 &0.38\\
\enddata
\tablecomments{Best fit parameters for circular motion in a plane
(with a small radial correction, a) for \neii\ emission and HCN(4-3)
emission in the Galactic center for different stellar mass distributions
holding the SMBH mass fixed at 4.2\ee{6}\,\msun and allowing it to be a 
free parameter.}
\tablenotetext{a}{The inclination is the angle between the sky plane and the plane of the disk.}
\tablenotetext{b}{The angle of line of nodes on the sky is the angle between
north and the intersection of the disk plane and the sky plane.}
\tablenotetext{c}{The radial velocity parameter (a) is the outward velocity
component as a fraction of the angular velocity (that is, the gravitational
circular velocity) multiplied by radius from the center in parsecs.
So at a distance of 1pc, and an angular velocity of $100$\,\kms,
$a=-0.05$\,pc implies and inward radial velocity of $5$\,\kms.}
\tablenotetext{d}{This is our ``goodness of fit" measurement (see Figure 5).
Note that for the unshifted spectrum this ratio is 0.15.}
\tablenotetext{e}{Consistent with the kinematics suggesting a black hole mass
that is too small, if we force the black hole to be 4.2\ee{6}\,\msun,
then we prefer no stars.
Naturally, we do not purport that there are no stars, nor that the
black hole mass is less than 4\ee{6}\,\msun, however
it is interesting to note what might cause these anomalies in the observational kinematics.
We speculate in Section 5.}
\tablenotetext{f}{After fitting a physical spiral with the parameters from
row 5 in this table, we fixed the inward velocity component to fit gas
flowing along this spiral to check that this parameter is not redundant
with the angle of line of nodes.
The best fit constraining $a=-0.27$\,pc did change $\Omega$ considerably, but
is a much worse fit kinematically.}
\end{deluxetable}

\clearpage

\end{document}